\documentclass[a4paper,10pt,twocolumn,showpacs,notitlepage,floatfix,nofootinbib,superscriptaddress,prd]{revtex4-2}
\usepackage{graphicx}
\usepackage{amsfonts}
\usepackage{amsmath}
\usepackage{hyperref}
\usepackage{float}
\usepackage{amssymb,amsbsy}
\usepackage{epstopdf}
\usepackage{color}
\usepackage{xcolor}
\usepackage{subfigure}
\usepackage[normalem]{ulem}
\usepackage{braket}
\usepackage{combelow}
\usepackage{ulem}
\usepackage{natbib}

\definecolor{goethe-blau}{cmyk}{1.0,0.2,0.0,0.4}
\definecolor{hellgrau}{cmyk}{0.04,0.04,0.05,0.02}
\definecolor{sandgrau}{cmyk}{0.12,0.09,0.13,0.0}
\definecolor{dunkelgrau}{cmyk}{0.25,0.25,0.30,0.75}
\definecolor{emo-rot}{cmyk}{0.04,1.0,0.8,0.07}
\definecolor{purple}{cmyk}{0.08,1.0,0.3,0.36}
\definecolor{senfgelb}{cmyk}{0.01,0.25,1.0,0.05}
\definecolor{gruen}{cmyk}{0.62,0.4,0.87,0.09}
\definecolor{magenta}{cmyk}{0.08,0.86,0.12,0.12}
\definecolor{orange}{cmyk}{0.0,0.7,1.0,0.04}
\definecolor{sonnengelb}{cmyk}{0.0,0.12,0.95,0.0}
\definecolor{helles-gruen}{cmyk}{0.4,0.17,0.81,0.07}
\definecolor{lichtblau}{cmyk}{0.8,0.0,0.06,0.04}

\newcommand{\Kn}{\mathrm{Kn}}
\renewcommand{\Re}{\mathrm{Re}^{-1}}
\newcommand{\bk}{\mathbf{k}}

\begin{document}

%\title{New reduction scheme for transient fluid dynamics from the Boltzmann equation}

\title{Inverse-Reynolds-Dominance approach to transient fluid dynamics}

\author{David Wagner}
	\email{dwagner@itp.uni-frankfurt.de}
    \affiliation{
Institut f\"ur Theoretische Physik, Johann Wolfgang Goethe-Universit\"at, Max-von-Laue-Straße 1, D-60438 Frankfurt am Main, Germany}%

\author{Andrea Palermo}
	\email{andrea.palermo@unifi.it}
    \affiliation{
Institut f\"ur Theoretische Physik, Johann Wolfgang Goethe-Universit\"at, Max-von-Laue-Straße 1, D-60438 Frankfurt am Main, Germany}%
    \affiliation{Dipartimento di Fisica e Astronomia, Universit\'a di Firenze and INFN Sezione di Firenze,
Via G. Sansone 1, Sesto Fiorentino I-50019, Firenze, Italy}

\author{Victor E. Ambru\cb{s}}
	\email[Corresponding author: ]{victor.ambrus@e-uvt.ro}
    \affiliation{
Institut f\"ur Theoretische Physik, Johann Wolfgang Goethe-Universit\"at, Max-von-Laue-Straße 1, D-60438 Frankfurt am Main, Germany}%
\affiliation{Department of Physics, West University of Timi\cb{s}oara, \\
Bd.~Vasile P\^arvan 4, Timi\cb{s}oara 300223, Romania}

% \begin{abstract}
% In this paper we present a novel formulation of second-order relativistic dissipative hydrodynamics from kinetic theory. Starting from the equations of motion for the irreducible moments of the single-particle distribution function and applying a power-counting scheme in Knudsen and inverse Reynolds numbers, we obtain relaxation equations for the bulk-viscous pressure, the diffusion current, and the shear-stress tensor, which can be systematically improved by including more moments. As a simplification compared to the DNMR formulation, no terms of second order in the Knudsen number appear, and no diagonalization of the linearized collision kernel is needed.
% This is \thumbsup.
% \end{abstract}

\begin{abstract}
We consider the evolution equations for the
bulk viscous pressure, diffusion current and 
shear tensor derived within second-order relativistic dissipative hydrodynamics from kinetic
theory. By matching the higher order moments
directly to the dissipative
quantities, all terms which are 
of second order in the Knudsen number $\Kn$ vanish, leaving only terms
of order $\mathcal{O}(\Re \, \Kn)$ and $\mathcal{O}({\rm Re}^{-2})$ in the 
relaxation equations, where $\Re$ is the inverse Reynolds number. We therefore refer to this scheme 
as the Inverse-Reynolds-Dominance (IReD) approach.
The remaining (non-vanishing) transport coefficients can be obtained 
exclusively in terms of the inverse of the
collision matrix. 
This procedure fixes unambiguously the relaxation times of the dissipative quantities, 
which are no longer related to the eigenvalues of
the inverse of the collision matrix. 
In particular, we find that the relaxation times corresponding to higher-order moments
grow as their order increases, thereby 
contradicting the \textit{separation of scales}
paradigm.
The formal (up to second order) equivalence with the standard DNMR 
approach is proven and the connection between the IReD transport coefficients 
and the usual DNMR ones is established.
\end{abstract}

\markboth{left head}{\today \; - \; 2021 v0.1}

\date{\today}
\maketitle
\section{Introduction}\label{sec:intro}
%The problem of 
Formulating a causal and stable framework for relativistic dissipative hydrodynamics has been a long-standing issue that has seen a series of improvements in the last decade \cite{Denicol.2012,Bemfica.2019,Kovtun.2019}. 
This problem is not merely academic, as dissipative fluid dynamics has been proven to be a powerful effective theory in relativistic systems, such as heavy-ion collisions \cite{GALE.2013,Romatschke.2019, Ryblewski.2011,Werner.2013}
and relativistic astrophysical processes \cite{Chabanov.2021,Turolla.1989, Tomei.2019}.

While the relativistic Euler equations describing the 
dynamics of the perfect fluid are unambiguously formulated,
their generalization to relativistic dissipative fluids proves 
to be a formidable problem. In the nonrelativisic case, the
leading-order contribution to the Chapman-Enskog expansion, 
i.e., the Navier-Stokes equations, yield a suitable theory 
for viscous hydrodynamics which has seen tremendous success
\cite{Landau.2014}. At this level, the dissipative quantities,
otherwise known as thermodynamic fluxes \cite{Rezzolla.2013},
are fixed by constitutive equations to the 
thermodynamic forces (expressed 
as gradients of the fluid properties),
thereby implying an instantaneous response 
and an infinite information propagation speed, thus violating causality \cite{Hiscock.1985,Hiscock.1987,Rezzolla.2013,Ambrus:2017keg}.
% and consequently leads to an unstable theory in reference frames where the fluid is moving \cite{Denicol.2008,Gavassino.2021}. 

An approach attracting much interest in recent years is to abandon the traditional (Landau or Eckart) matching conditions, by which the energy and particle number density of the system are equated to their fictitious local-equilibrium counterparts. In contrast, general matching conditions can be exploited in the frame of a first order-like theory closely resembling the Navier-Stokes formulation in a way that guarantees causality and stability \cite{Kovtun.2019, Bemfica.2019, Bemfica.2021, Rocha.2021}.

In this paper, we focus on the more traditional approach 
of formulating a causal and stable theory of dissipative
hydrodynamics in the form of relaxation equations for the 
dissipative quantities appearing in the particle current and stress-energy tensor
decompositions, namely the bulk-viscous pressure $\Pi$, the 
particle diffusion current $n^\mu$, and the shear-stress tensor $\pi^{\mu\nu}$. 
Such second-order theories introduce 
relaxation times governing the response of the 
dissipative quantities with respect to changes in the 
fluid properties 
(e.g., pressure $P$, ratio $\alpha = \mu / T$ between the chemical potential $\mu$ and temperature $T$,
and four-velocity $u^\mu$). 
This procedure sets finite relaxation time scales 
of the approach towards
the corresponding asymptotic Navier-Stokes limits, 
thereby rendering the formulation causal 
\cite{Denicol.2011}.

Naturally, due to the microscopic nature of the coefficients involved in second-order theories, an underlying formulation has to be provided. 
Most works employ
kinetic theory, since it provides a suitable limit of quantum field theories in the semiclassical limit \cite{Groot.1980}. From a thermodynamical perspective, the entropy current describing the entropy flow in second-order hydrodynamics exhibits second-order terms, which are in principle calculable from kinetic theory \cite{Israel.1979,Groot.1980,El:2008yy} or can be postulated within the frame of extended irreversible thermodynamics \cite{Israel.1976,Hiscock.1983,Jou.1985,Muronga.2004,Denicol.2009,Gavassino.2021_antonelli}.

Even though the second-order formalism by Müller, Israel and Stewart \cite{Muller.1967,Israel.1979} has long been the most widely used second-order theory, its equations of motion were obtained by employing a non-orthogonal momentum space basis which cannot be used to systematically account for all second-order terms, leading to inaccurate expressions for the transport coefficients.
This issue can be resolved by considering the exact equations of motion for the irreducible moments of the Boltzmann equation, as has been done in the celebrated
DNMR formulation of relativistic dissipative hydrodynamics \cite{Denicol.2012,Molnar.2014,Denicol.2018,Denicol.2019}. In addition, this formulation provided a way to improve the truncation by increasing the number of moments considered for a given tensor-rank, thereby moving from the (lowest-order) 14-moment approximation to 23 moments and beyond. 

In the DNMR formulation \cite{Denicol.2012,Denicol.2021}, the main idea consists of obtaining a system of relaxation equations for the eigenmodes of the linearized collision kernel, which can then be related to the irreducible moments and thus to the dissipative quantities. 
The evolution equations for the dissipative quantities obtained 
in this procedure read
\begin{subequations} \label{eq:DNMR_eqs}
\begin{align}
 \tau_\Pi \dot{\Pi} +\Pi =& -\zeta \theta + \mathcal{J} + 
 \mathcal{K} + \mathcal{R}\;,\\
 \tau_n \dot{n}^{\langle \mu \rangle} +n^\mu =&\, \kappa I^\mu +\mathcal{J}^\mu +
 \mathcal{K}^\mu + \mathcal{R}^{\mu}\;,\\
 \tau_\pi \dot{\pi}^{\langle \mu\nu \rangle} +\pi^{\mu\nu} =&\, 
 2\eta \sigma^{\mu\nu}+\mathcal{J}^{\mu\nu} + \mathcal{K}^{\mu\nu} + 
 \mathcal{R}^{\mu\nu}\;,
\end{align}
\end{subequations}
where $\tau_\Pi$, $\tau_n$ and $\tau_\pi$ are the relaxation 
time corresponding to $ \Pi$, $n^\mu$ and $\pi^{\mu\nu}$, respectively, 
while $\zeta$, $\kappa$ and $\eta$ are the bulk viscosity, diffusivity and shear viscosity coefficients, constituting 
the so-called first-order (Navier-Stokes) transport coefficients. 
In this procedure, the relaxation equations for the dissipative quantities 
are derived on the basis of a hierachical truncation 
with respect to the inverse Reynolds number $\Re$, related 
to the magnitude of the dissipative quantities 
($\Re \sim |\Pi| / P_0,\, |n^\mu| / n_0,\, |\pi^{\mu\nu}| / P_0$), as well as to the
Knudsen number $\Kn$, which can be related 
to the magnitude of gradients
($\Kn \sim \ell \nabla A / A_0$, where $\ell$ is a
characteristic microscopic scale and $A$ is a fluid property) 
or to the microscopic mean free path ($\Kn \sim \tau_\Pi / L,\,
\tau_n / L,\, \tau_\pi / L$, where $L$ is a macroscopic length scale). 
By the above definitions, the second-order terms 
$\mathcal{J}^{\mu_1 \cdots \mu_\ell}$,
$\mathcal{K}^{\mu_1 \cdots \mu_\ell}$, 
and $\mathcal{R}^{\mu_1 \cdots \mu_\ell}$ collect 
all contributions that are of orders $\mathcal{O}(\Re\ \Kn)$, 
$\mathcal{O}(\Kn^2)$ and $\mathcal{O}({\rm Re}^{-2})$, respectively.
All third order terms with respect to $\Kn$ and $\Re$, i.e., the terms of orders $\mathcal{O}(\mathrm{Re}^{-3})$, $\mathcal{O}(\mathrm{Re}^{-2} \Kn)$,
$\mathcal{O}(\Re \Kn^2)$, and $\mathcal{O}(\Kn^3)$, are neglected. It is thus implied that, besides the requirement that both $\Re$ and $\Kn$ are small, the above power-counting scheme also assumes that they are of the same magnitude, i.e., $\Kn\sim\Re$. It should be noted that in general the magnitudes of $\Kn$ and $\Re$ need not be the same, hence the regimes where only one of these quantities is small and the other one is large lie outside the domain of applicability of second-order hydrodynamics. Such regimes may be probed using third-order hydrodynamics \cite{Jaiswal:2013vta,Grozdanov:2015kqa} or directly kinetic theory \cite{Xu:2004mz,Bouras:2009nn,Florkowski:2013lya,Florkowski:2014sfa,Denicol:2012vq,Denicol:2014xca,Ambrus:2017keg}, however in this paper we focus only on the second-order theory.

The $\mathcal{R}^{\mu_1 \cdots \mu_\ell}$ terms, derived 
in Ref.~\cite{Molnar.2014}, arise 
due to quadratic terms appearing in the collision term. 
These terms will not be discussed further in this work.
The $\mathcal{K}^{\mu_1 \cdots \mu_\ell}$ terms 
involve quadratic terms in the first order gradients of the 
flow properties 
(e.g., $\sigma^{\lambda \langle \mu} \sigma^{ \nu\rangle}{}_\lambda$)
or second order gradients (e.g., 
$\Delta^\mu_\lambda \nabla_\nu \sigma^{\lambda \nu}$). 
Their transport coefficients were 
derived in Ref.~\cite{Molnar.2014}, however they are 
usually disregarded because they give rise to parabolic equations \cite{Denicol:2012vq}.
On the other hand, the terms 
in $\mathcal{J}^{\mu_1 \cdots \mu_\ell}$ are hyperbolic in nature 
and are fully compatible with special relativity.

In this paper we show that it is possible to formulate a theory of dissipative relativistic hydrodynamics setting the non-causal contribution $\mathcal{K}^{\mu_1\cdots\mu_\ell}$ to zero by construction. The basis of our analysis is 
the asymptotic matching scheme proposed in 
Ref.~\cite{Denicol:2012vq} in the context of multiple 
dynamical moments, as well as in 
Ref.~\cite{Jan} for the case of multicomponent fluids. 
The scheme finds its non-relativistic analogue in 
the work of Struchtrup \cite{Struchtrup.2004}, and it is sometimes called \emph{order of magnitude approach}. 

Except in the case of the lowest-order truncation, the transport coefficients and the relaxation times obtained in this scheme are different compared to those obtained in DNMR. The two theories thus 
seem to yield, in general, different equations.
In this paper we establish
the connection between the two schemes
and show that they are equivalent up to second order in $\Kn$ and $\Re$.
By consistently using the matching conditions to express thermodynamic forces in terms of dissipative quantities, we show that all terms contained in $\mathcal{K}^{\mu_1\cdots\mu_\ell}$ in DNMR can be reabsorbed into the transport coefficients in $\mathcal{J}^{\mu_1\cdots\mu_\ell}$ 
and the relaxation times, 
 thus modifying
the usual DNMR transport coefficients. 
We therefore call our approach the
\emph{Inverse-Reynolds-Dominance} (IReD) approach, as it consists, effectively, in replacing $\mathcal{O}(\Kn^2)$ terms in favour of $\mathcal{O}(\Re\Kn)$, making the inverse Reynolds number ``dominant" over the Knudsen number. 
The IReD equations are formally equivalent to the DNMR ones. We will show this by analytically establishing the connection between the transport coefficients appearing in the two formulations.

The outline of this paper is as follows.
In Sec.~\ref{sec:DNMR}, we review the DNMR formalism introduced in 
Ref.~\cite{Denicol.2012}, while in Sec.~\ref{sec:IReD} we discuss 
the IReD 
scheme, leading to 
vanishing $\mathcal{K}^{\mu_1\cdots \mu_\ell}$ 
terms \cite{Jan}. 
Section~\ref{sec:connection} addresses the connection between the 
transport coefficients arising in the IReD approach 
compared to the DNMR ones (technical details are relegated to 
Appendix~\ref{app:useful formulae}). 
Section~\ref{sec:connection2} discusses
the connection between the approach introduced in
Ref.~\cite{Denicol:2012vq} for the case of 23 dynamical
moments and our proposed IReD approach.
In Sec. \ref{sec:UR} 
we list the explicit values for the transport coefficients in 
the limit of an ultrarelativistic ideal gas of hard spheres,
demonstrating the convergence of the method when including
higher-order moments. The general expressions for the transport coefficients in the IReD approach are summarized in Appendix~\ref{app:tcoeffs}. Section~\ref{sec:conc} concludes this paper.
Throughout this paper, we use Planck units ($c = \hbar = k_B = 1$) and 
the $(+,-,-,-)$ metric convention. 
Our analysis is restricted to second order with respect to $\Kn$ and $\Re$ and we work under the assumption that $\Kn \sim \Re$.

\section{DNMR approach}\label{sec:DNMR}

In this section, we review the DNMR formalism
introduced in Ref.~\cite{Denicol.2012}.
The starting point of the analysis is the Boltzmann equation,
\begin{equation}
 k^\mu \partial_\mu f_\bk = C[f]\;,\label{eq:boltz}
\end{equation}
where $f_\bk \equiv f_\bk(x)$ is the one-particle distribution
function, $k^\mu = (k^0, \mathbf{k})$ is the on-shell four-momentum
($k^2 = (k^0)^2 - \mathbf{k}^2 = m^2$), while $C[f]$ is 
the collision term.
By the $H$-theorem \cite{Groot.1980,Cercignani.2002,LandauX}, $C[f]$ acts by 
drawing the system towards local thermodynamic equilibrium,
described by the equilibrium distribution $f_{0\bk}$.

The deviation from equilibrium $\delta f_\bk = f_\bk - f_{0\bk}$ can be characterized in terms 
of its irreducible moments $\rho^{\mu_1 \cdots \mu_\ell}_r$, defined as
\begin{equation}
 \rho^{\mu_1 \cdots \mu_\ell}_r = 
 \int dK \, E_\bk k^{\langle \mu_1} \cdots k^{\mu_\ell \rangle} 
 \delta f_\bk\;,
 \label{eq:rho_def}
\end{equation}
where $dK = g d^3k / [(2\pi)^3 k^0]$ is the Lorentz-invariant integration measure 
($g$ is the number of internal degrees of freedom), while
$A^{\langle \mu_1 \cdots \mu_\ell\rangle} = \Delta^{\mu_1 \cdots \mu_\ell}_{\nu_1 \cdots \nu_\ell} A^{\nu_1 \cdots \nu_\ell}$ is the symmetrized and (for $\ell > 1$) traceless projection of the tensor $A^{\mu_1 \cdots \mu_\ell}$ with respect to the 
fluid four-velocity $u^\mu$. In particular, the $r = 0$ moments can be related to 
the bulk pressure $\Pi$, 
diffusion current $n^\mu$ and shear stress $\pi^{\mu\nu}$ as follows:
\begin{equation}\label{eq:zeroth moments}
 \rho_0 = -\frac{3}{m^2} \Pi\;, \qquad 
 \rho_0^\mu = n^\mu\;, \qquad 
 \rho_0^{\mu\nu} = \pi^{\mu\nu}\;.
\end{equation}
In the Landau frame, the charge current $N^\mu$ and 
stress-energy tensor $T^{\mu\nu}$ admit the following decomposition:
\begin{equation}
 N^\mu = nu^\mu + n^\mu\;, \qquad 
 T^{\mu\nu} = \varepsilon u^\mu u^\nu - (P + \Pi) \Delta^{\mu\nu} 
 + \pi^{\mu\nu}\;,
\end{equation}
where $\Delta^{\mu\nu} = g^{\mu\nu} - u^\mu u^\nu$.
Since the particle-number density $n$ and energy density $\varepsilon$ 
are equal to their fictitious equilibrium values ($n = n_0$, $\varepsilon = \varepsilon_0$), the moments $\rho_1 = \delta n$ and $\rho_2 = \delta \varepsilon$ 
both vanish. In addition, the heat flow $W^\mu = \Delta^\mu_\nu u_\lambda T^{\nu\lambda} = \rho_1^\mu$ also vanishes by the Landau matching condition, $T^\mu_\nu u^\nu = \varepsilon u^\mu$. Summarizing, in the Landau frame the following moments are automatically 
zero:
\begin{equation}
    \rho_1 = \rho_2 = \rho_1^\mu = 0\;.
    \label{eq:rho_Landau}
\end{equation}

Starting from the Boltzmann equation (2) and defining $\nabla^\mu=\Delta^\mu_\nu\partial^\nu$ and $\dot{f} = Df = u\cdot\partial f$ for an arbitrary function f, the equations of motion for the irreducible moments 
$\rho_r$, $\rho^\mu_r$ and $\rho^{\mu\nu}_r$ can be 
derived as shown in Ref.~\cite{Denicol.2012}, leading to
\begin{widetext}
\begin{subequations}\label{eq:rhodot_all}
\begin{align}
 \dot{\rho}_r - C_{r-1} =&\, \alpha_r^{(0)} \theta -
 \frac{G_{2r}}{D_{20}} \Pi \theta + 
 \frac{G_{2r}}{D_{20}} \pi^{\mu\nu} \sigma_{\mu\nu} + 
 \frac{G_{3r}}{D_{20}} \partial_\mu n^\mu
 + (r - 1) \rho^{\mu\nu}_{r-2} \sigma_{\mu\nu} + 
 r \rho^\mu_{r-1} \dot{u}_\mu - \nabla_\mu \rho^\mu_{r-1}\nonumber\\
 &- \frac{1}{3}\left[(r+2) \rho_r - (r -1)m^2 \rho_{r-2}\right]
 \theta\;, \label{eq:rhodot_l0}\\
 \dot{\rho}^{\langle\mu\rangle}_r - C_{r-1}^{\langle \mu \rangle} =&\, \alpha_r^{(1)} I^\mu + 
 \rho^\nu_r \omega^\mu{}_\nu + 
 \frac{1}{3}[(r-1) m^2 \rho^\mu_{r-2} - (r+ 3) \rho^\mu_r] \theta 
 - \Delta^\mu_\lambda \nabla_\nu \rho^{\lambda \nu}_{r-1} 
 + r \rho^{\mu\nu}_{r-1} \dot{u}_\nu \nonumber\\
 &+ \frac{1}{5} \left[(2r-2) m^2 \rho^\nu_{r-2} - (2r+3) \rho^\nu_r\right] \sigma^\mu_\nu 
 + \frac{1}{3}\left[m^2 r \rho_{r-1} - (r+3) \rho_{r+1}\right] \dot{u}^\mu \nonumber\\
 &+ \frac{\beta J_{r+2,1}}{\varepsilon + P} 
 (\Pi \dot{u}^\mu - \nabla^\mu \Pi + \Delta^\mu_\nu \partial_\lambda \pi^{\lambda \nu}) 
 -\frac{1}{3} \nabla^\mu (m^2 \rho_{r-1} - \rho_{r+1}) 
 + (r -1 ) \rho^{\mu\nu\lambda}_{r-2} \sigma_{\lambda\nu}\;,
 \label{eq:rhodot_l1} \\
 \dot{\rho}^{\langle \mu\nu\rangle}_r - C^{\langle \mu\nu \rangle}_{r-1} =& \,
 2 \alpha^{(2)}_r \sigma^{\mu\nu} - \frac{2}{7} \left[(2r+5) \rho^{\lambda\langle \mu}_r - 2m^2(r-1) \rho^{\lambda\langle \mu}_{r-2}\right] \sigma^{\nu\rangle}_\lambda + 
 2\rho^{\lambda\langle \mu}_r \omega^{\nu\rangle}{}_\lambda \nonumber\\
 &+ \frac{2}{15}[(r+4)\rho_{r+2} - (2r+3) m^2 \rho_r + 
 (r - 1)m^4 \rho_{r-2}] \sigma^{\mu\nu} + 
 \frac{2}{5} \nabla^{\langle \mu} (\rho^{\nu \rangle}_{r+1} - m^2 \rho^{\nu\rangle}_{r-1}) \nonumber\\
 &- \frac{2}{5} \left[(r+5) \rho^{\langle \mu}_{r+1} - r m^2 \rho^{\langle \mu}_{r-1}\right] \dot{u}^{\nu \rangle} 
 - \frac{1}{3} \left[(r + 4) \rho^{\mu\nu}_r - m^2 (r-1) \rho^{\mu\nu}_{r-2}\right] \theta 
 \nonumber\\
 & + (r - 1) \rho^{\mu\nu\lambda\rho}_{r-2} \sigma_{\lambda \rho} - \Delta^{\mu\nu}_{\alpha\beta} 
 \nabla_\lambda \rho^{\alpha\beta\lambda}_{r-1} + 
 r \rho^{\mu\nu\lambda}_{r-1} \dot{u}_\lambda\;,
 \label{eq:rhodot_l2}
\end{align}
\end{subequations}
\end{widetext}
where $I^\mu =\nabla^\mu \alpha$. Furthermore, $\sigma_{\mu\nu}=\nabla_{\langle\mu} u_{\nu\rangle}$, $\omega_{\mu\nu}=\frac{1}{2} (\partial_\mu u_\nu-\partial_\nu u_\mu)$ and $\theta=\partial_\mu u^\mu$ denote the shear tensor, vorticity tensor 
and expansion scalar, respectively, while
$C^{\langle \mu_1 \cdots \mu_\ell\rangle}_{r-1}$ represents an irreducible moment of tensor-rank $\ell$
of the collision term, defined in analogy to Eq.~\eqref{eq:rho_def}:
\begin{equation}
 C^{\langle \mu_1 \cdots \mu_\ell \rangle}_r = 
 \int dK \, E_\bk^r k^{\langle \mu_1} \cdots k^{\mu_\ell \rangle} 
 C[f]\;.
 \label{eq:Cr_def}
\end{equation}
Furthermore, $G_{nm} = J_{n0} J_{m0} - J_{n-1,0} J_{m+1,0}$, $D_{nq} = J_{n+1,q} J_{n-1,q} - J_{nq}^2$, while $J_{nq} = (\partial I_{nq} / \partial \alpha)_\beta$ represents the 
derivative of $I_{nq}$ with respect to $\alpha = \mu / T$ at constant inverse temperature $\beta$, with
\begin{equation}
    I_{nq} = \frac{1}{(2q+1)!!} \int dK\, E_\bk^{n-2q} (-\Delta^{\alpha\beta} k_\alpha k_\beta)^q f_{0\bk}\;.\label{eq:I_def}
\end{equation}
The first terms appearing on the right-hand sides of Eqs.~\eqref{eq:rhodot_all} are given as
\begin{align}
 \alpha_r^{(0)} =& (1 - r) I_{r1} - I_{r0} - \frac{1}{D_{20}} 
 [G_{2r}(\varepsilon + P) - G_{3r} n]\;,\nonumber\\
 \alpha_r^{(1)} =& J_{r+1,1} - \frac{n}{\varepsilon + P} J_{r+2,1}\;,\nonumber\\
 \alpha_r^{(2)} =& I_{r+2,1} + (r - 1) I_{r+2,2}\;.\label{eq:alpha}
\end{align}

The relations \eqref{eq:rhodot_all} constitute a system of infinitely many coupled 
equations, where 
the unknowns are the irreducible moments $\rho^{\mu_1 \cdots \mu_\ell}_r$. In order 
to extract from here the equations of motion for the dissipative quantities 
$\Pi$, $n^\mu$ and $\pi^{\mu\nu}$, the collision term $C_r^{\langle \mu_1 \cdots \mu_\ell \rangle}$ must 
be expressed in terms of $\rho^{\mu_1 \cdots \mu_\ell}_r$. 
This can be achieved by introducing a decomposition of $\delta f_\bk$ with respect to
the irreducible moments,
\begin{equation}
 \delta f_\bk =
 f_{0\bk} \tilde{f}_{0\bk}
 \sum_{\ell=0}^\infty \sum_{n=0}^{N_\ell} \mathcal{H}_{\bk n}^{(\ell)} \rho_{n}^{\mu_1\cdots \mu_\ell}k_{\langle\mu_1}\cdots k_{\mu_\ell\rangle},
 \label{eq:df}
\end{equation}
where $\tilde{f}_{0\bk} = 1 - a f_{0\bk}$  ($a = 1$ for fermions, 
$-1$ for bosons and $0$ for classical particles) and $N_\ell$ represent 
truncation orders which in principle should be taken to infinity. 
The functions $\mathcal{H}^{(\ell)}_{\bk n}$ are polynomials of order 
$N_\ell$ in $E_\bk$ and are defined such that 
Eq.~\eqref{eq:rho_def} holds exactly for $0 \le n \le N_\ell$ \cite{Denicol.2012}.

Ignoring quadratic or higher-order terms in deviations from equilibrium, 
the collision term can be represented (to linear order) as 
\begin{subequations} \label{eq:A_def}
\begin{align}
 C_{r-1} =&  - \sum_{n = 0,\neq 1,2}^{N_0} \mathcal{A}^{(0)}_{rn} \rho_n\;,
\\
 C^{\langle \mu \rangle}_{r-1} =& 
 - \sum_{n = 0,\neq 1}^{N_1} \mathcal{A}^{(1)}_{rn} \rho^\mu_n\;,
\\
 C^{\langle \mu \nu \rangle}_{r-1} =& 
 - \sum_{n = 0}^{N_2} \mathcal{A}^{(2)}_{rn} \rho^{\mu\nu}_n\;,
\end{align}
\end{subequations}
where $\mathcal{A}^{(\ell)}_{rn}$ can be interpreted as the collision matrix. 
The sums appearing above skip the moments which vanish due to the Landau matching,
as shown in Eq.~\eqref{eq:rho_Landau}.

The final step is to relate the irreducible moments $\rho^{\mu_1 \cdots \mu_\ell}_{r \neq 0}$ 
to those of order $r = 0$. This is the branching point between the DNMR approach 
and the IReD approach presented in Sec.~\ref{sec:IReD}.
In the DNMR approach, the basis of this construction is to seek a diagonalization 
of $\mathcal{A}^{(\ell)}_{rn}$ ensured by the matrix of eigenvectors $\Omega^{(\ell)}_{rn}$, such that
\begin{equation}
 (\Omega^{(\ell)})^{-1} \mathcal{A}^{(\ell)} \Omega^{(\ell)} = 
 {\rm diag}(\chi_0^{(\ell)},\; \chi_1^{(\ell)},\; \cdots)\;,
 \label{eq:DNMR_eigen}
\end{equation}
where the columns of the diagonalization matrix $\Omega^{(\ell)}$ 
are chosen such that the eigenvalues $\chi_r^{(\ell)}$ appear in 
ascending order,
\begin{equation}
 \chi^{(\ell)}_0 \le \chi^{(\ell)}_1 \le \cdots\;.
\end{equation}
With the above convention,
it is possible to enforce a {\it separation of scales} 
by which only the eigenvectors 
\begin{equation}
 X_0^{\mu_1 \cdots \mu_\ell} = \sum_{j = 0}^{N_\ell} 
 (\Omega^{(\ell)})^{-1}_{0j} \rho_j^{\mu_1 \cdots \mu_\ell}
 \label{eq:DNMR_X0}
\end{equation}
corresponding to the slowest scale $\chi_0^{(\ell)}$ remain in the transient regime 
(the normalization of $\Omega^{(\ell)}_{ij}$ is such that 
$\Omega^{(\ell)}_{00} = 1$). 
% \begin{align}
%  \dot{X}_0 + \chi_0^{(0)} X_0 \simeq& \beta_0^{(0)} \theta,\nonumber\\
%  \dot{X}_0^\mu + \chi_0^{(1)} X_0^\mu \simeq& \beta_0^{(1)} I^\mu,\nonumber\\
%  \dot{X}_0^{\mu\nu} + \chi_0^{(2)} X_0^{\mu\nu} \simeq& \beta_0^{(2)} \sigma^{\mu\nu},
% \end{align}
The eigenvectors $X_{r > 0}^{\mu_1 \cdots \mu_\ell}$, corresponding to larger eigenvalues 
$\chi^{(\ell)}_{r>0}$, are approximated by their asymptotic (Navier-Stokes)
values
\begin{equation}
 X_{r>0} \simeq \frac{\beta_r^{(0)}}{\chi^{(0)}_r} \theta\;,\;\;
 X^\mu_{r>0} \simeq \frac{\beta_r^{(1)}}{\chi^{(1)}_r} I^\mu\;,\;\;
 X^{\mu\nu}_{r>0} \simeq \frac{\beta_r^{(2)}}{\chi^{(2)}_r} \sigma^{\mu\nu}\;,
\end{equation}
where
\begin{subequations}
\begin{align}
 \beta_i^{(0)} =& \sum_{j = 0,\neq 1,2}^{N_0} (\Omega^{(0)})^{-1}_{ij} \alpha^{(0)}_j\;,\\
 \beta_i^{(1)} =& \sum_{j = 0,\neq 1}^{N_1} (\Omega^{(1)})^{-1}_{ij} \alpha^{(1)}_j\;,\\
 \beta_i^{(2)} =& 2\sum_{j = 0}^{N_2} (\Omega^{(2)})^{-1}_{ij} \alpha^{(2)}_j\;.
\end{align}
\end{subequations}
By this approximation, the irreducible moments $\rho^{\mu_1 \cdots \mu_\ell}_r = \sum_{n = 0}^{N_\ell} \Omega^{(\ell)}_{rn} X_n^{\mu_1 \cdots \mu_\ell}$ take the following 
form,
\begin{subequations}\label{eq:DNMR_matching}
\begin{align}
 \rho_i \simeq& -\frac{3}{m^2}[\Omega^{(0)}_{i0} \Pi 
 - (\zeta_i - \Omega^{(0)}_{i0} \zeta) \theta]\;, \\
 \rho^\mu_i \simeq&\, \Omega^{(1)}_{i0} n^\mu
 + (\kappa_i - \Omega^{(1)}_{i0} \kappa) I^\mu\;,\\
 \rho^{\mu\nu}_i \simeq&\, \Omega^{(2)}_{i0} \pi^{\mu\nu} + 2(\eta_i - \Omega^{(2)}_{i0} \eta) \sigma^{\mu\nu},\label{eq:DNMR_matching_2}\\
 \rho^{\mu\nu \lambda \cdots}_i \simeq&\, \mathcal{O}(\Kn^2, \Kn\ \Re)\;,
\end{align}
\end{subequations}
where the property $X_0^{\mu_1 \cdots \mu_\ell} = \rho_0^{\mu_1 \cdots\mu_\ell} - \sum_{n > 0}^{N_\ell} \Omega^{(\ell)}{rn} X_n^{\mu_1 \cdots \mu\ell}$ was employed. In the above, the first-order transport coefficients
$\zeta_i$, $\kappa_i$ and $\eta_i$ are computed via
\begin{subequations}\label{eq:NS_coeff}
\begin{align}
 \zeta_n =& \frac{m^2}{3} \sum_{r = 0, \neq 1,2}^{N_0} 
 \tau^{(0)}_{nr} \alpha^{(0)}_r\;, \\
 \kappa_n =& \sum_{r = 0, \neq 1}^{N_1} 
 \tau^{(1)}_{nr} \alpha^{(1)}_r\;, \\
 \eta_n =& \sum_{r = 0}^{N_2} 
 \tau^{(2)}_{nr} \alpha^{(2)}_r\;,
\end{align}
\end{subequations}
with $\zeta = \zeta_0$, $\kappa = \kappa_0$ and $\eta = \eta_0$.
The inverse collision matrix $\tau^{(\ell)}_{rn}$ appearing above
satisfies
\begin{equation}
 \tau^{(\ell)}_{rn} = (\mathcal{A}^{(\ell)})^{-1}_{rn} = 
 \sum_{m = 0}^{N_\ell} \Omega^{(\ell)}_{rm}
 \frac{1}{\chi^{(\ell)}_m} (\Omega^{(\ell)})^{-1}_{mn}\;.
 \label{eq:DNMR_tau_def}
\end{equation}

In what concerns the moments of negative 
order $\rho^{\mu_1 \cdots \mu_\ell}_{-r}$ (with $r > 0$), 
they can also be related to the dissipative quantities via
\begin{equation}
 \rho^{\mu_1 \cdots \mu_\ell}_{-r} = \sum_{n = 0}^{N_\ell} \mathcal{F}^{(\ell)}_{rn} 
 \rho_n^{\mu_1 \cdots \mu_\ell}\;,
\end{equation}
where the functions $\mathcal{F}^{(\ell)}_{rn}$ are defined as
\begin{equation}
 \mathcal{F}^{(\ell)}_{rn} = \frac{\ell!}{(2\ell + 1)!!} \int dK\,
 f_{0\bk} \tilde{f}_{0\bk} E_\bk^{-r} \mathcal{H}^{(\ell)}_{\bk n}
 (\Delta^{\alpha\beta} k_\alpha k_\beta)^\ell\;,
 \label{eq:DNMR_F}
\end{equation}
which follows after introducing Eq.~\eqref{eq:df} into Eq.~\eqref{eq:rho_def}.
Using now the asymptotic matching in Eqs.~\eqref{eq:DNMR_matching}, we arrive at
\begin{subequations}\label{eq:DNMR_rho_neg}
\begin{align}
 \rho_{-r} =& -\frac{3}{m^2} (\gamma_r^{(0)} \Pi - \hat{\gamma}_r^{(0)} \theta)\;,\\
 \rho^\mu_{-r} =& \gamma_r^{(1)} n^\mu + \hat{\gamma}_r^{(1)} I^\mu\;,\\
 \rho^{\mu\nu}_{-r} =& \gamma_r^{(2)} \pi^{\mu\nu} + 2\hat{\gamma}_r^{(2)} \sigma^{\mu\nu}\;.
\end{align}
\end{subequations}
The coefficients $\gamma_r^{(\ell)}$ and $\hat{\gamma}_r^{(\ell)}$ can be computed 
using the functions $\mathcal{F}^{(\ell)}_{rn}$,
\begin{align}
 \gamma_r^{(0)} =& \sum_{n = 0,\neq 1,2}^{N_0} 
 \mathcal{F}^{(0)}_{rn} \Omega^{(0)}_{n0}\;, & 
 \hat{\gamma}_r^{(0)} =& \sum_{n = 0,\neq 1,2}^{N_0} 
 \mathcal{F}^{(0)}_{rn} (\zeta_n - \Omega^{(0)}_{n0} \zeta)\;,\nonumber\\
 \gamma_r^{(1)} =& \sum_{n = 0,\neq 1}^{N_1} 
 \mathcal{F}^{(1)}_{rn} \Omega^{(1)}_{n0}\;, & 
 \hat{\gamma}_r^{(1)} =& \sum_{n = 0,\neq 1}^{N_1} 
 \mathcal{F}^{(1)}_{rn} (\kappa_n - \Omega^{(1)}_{n0} \kappa)\;,\nonumber\\
 \gamma_r^{(2)} =& \sum_{n = 0}^{N_2}
 \mathcal{F}^{(2)}_{rn} \Omega^{(2)}_{n0}\,, & 
 \hat{\gamma}_r^{(2)} =& \sum_{n = 0}^{N_2}
 \mathcal{F}^{(2)}_{rn} (\eta_n - \Omega^{(2)}_{n0} \eta)\;.
\end{align}
At this point, we remark that in the DNMR approach \cite{Denicol.2012} and in 
later papers \cite{Molnar.2014}, the terms $\hat{\gamma}^{(\ell)}_r$ are neglected, such that the 
$\mathcal{O}(\Kn)$ contributions to $\rho_{-r}^{\mu_1 \cdots \mu_\ell}$ that should
later appear in the $\mathcal{K}^{\mu_1 \cdots \mu_\ell}$ terms are disregarded completely \cite{Molnar.2014}. In order to conform with the DNMR notation and still stay accurate 
at first order with respect to both $\Kn$ and $\Re$, the 
coefficient $\gamma^{(\ell)}_r$ should be replaced by
\begin{subequations} \label{eq:DNMR_gamma_bar}
\begin{align}
 \bar{\gamma}^{(0)}_r =& \gamma^{(0)}_r + \frac{1}{\zeta} \hat{\gamma}^{(0)}_r 
 = \sum_{n = 0,\neq 1,2}^{N_0} \mathcal{F}^{(0)}_{rn} \mathcal{C}^{(0)}_n\;,\\
 \bar{\gamma}^{(1)}_r =& \gamma^{(1)}_r + \frac{1}{\kappa} \hat{\gamma}^{(1)}_r 
 = \sum_{n = 0,\neq 1}^{N_1} \mathcal{F}^{(1)}_{rn} \mathcal{C}^{(1)}_n\;,\\
 \bar{\gamma}^{(2)}_r =& \gamma^{(2)}_r + \frac{1}{\eta} \hat{\gamma}^{(2)}_r 
 = \sum_{n = 0}^{N_2} \mathcal{F}^{(2)}_{rn} \mathcal{C}^{(2)}_n\;,
\end{align}
\end{subequations}
where we introduced the notation (also to be used in the following section)
\begin{equation}
 \mathcal{C}^{(0)}_n = \frac{\zeta_n}{\zeta_0}\;, \qquad 
 \mathcal{C}^{(1)}_n = \frac{\kappa_n}{\kappa_0}\;, \qquad 
 \mathcal{C}^{(2)}_n = \frac{\eta_n}{\eta_0}\;.
 \label{eq:C_def}
\end{equation}
The same quantities are denoted in Ref.~\cite{Jan} 
by $\bar{\zeta}_n = \mathcal{C}^{(0)}_n$, $\bar{\kappa}_n = \mathcal{C}^{(1)}_n$ and 
$\bar{\eta}_n = \mathcal{C}^{(2)}_n$.
With the above convention, Eqs.~\eqref{eq:DNMR_rho_neg} becomes
\begin{align}
 \rho_{-r} =& -\frac{3}{m^2} \bar{\gamma}_r^{(0)} \Pi\;, &
 \rho^\mu_{-r} =& \bar{\gamma}_r^{(1)} n^\mu\;, &
 \rho^{\mu\nu}_{-r} =& \bar{\gamma}_r^{(2)} \pi^{\mu\nu}\;,
 \label{eq:C_rho_neg}
\end{align}
which is similar, but not identical, to Eq.~(67) in Ref.~\cite{Denicol.2012}.

Finally, the evolution equations \eqref{eq:DNMR_eqs} for $\Pi$, $n^\mu$ and $\pi^{\mu\nu}$ can be obtained 
by multiplying Eqs.~\eqref{eq:rhodot_all} by $\tau^{(\ell)}_{0r}$ and summing over $r$.
% In what follows, we use a hat $\hat{\phantom{\eta}}$ in order to distinguish 
% the transport coefficients appearing in the DNMR formulation from those
% appearing in the next section.
The relaxation times $\tau_\Pi$, $\tau_n$ and $\tau_\pi$ 
are given by the inverse of the smallest eigenvalues $\chi^{(\ell)}_0$ of the 
collision matrices $\mathcal{A}^{(\ell)}_{rn}$ [see Eq.~\eqref{eq:DNMR_eigen}]:
\begin{subequations} \label{eq:DNMR_rtimes}
\begin{align}
 \tau_\Pi =& \frac{1}{\chi^{(0)}_0} = \sum_{r = 0,\neq 1,2}^{N_0} 
 \tau^{(0)}_{0r} \Omega^{(0)}_{r0}\;,\\
 \tau_n =& \frac{1}{\chi^{(1)}_0} = \sum_{r = 0,\neq 1}^{N_1} 
 \tau^{(1)}_{0r} \Omega^{(1)}_{r0}\;,\\
 \tau_\pi =& \frac{1}{\chi^{(2)}_0} = \sum_{r = 0}^{N_2} 
 \tau^{(2)}_{0r} \Omega^{(2)}_{r0}\;,
 \label{eq:DNMR_rtimes_2}
\end{align}
\end{subequations}
where we remind that the normalization of $\Omega^{(\ell)}_{rn}$ is such that 
$\Omega^{(\ell)}_{00} = 1$.
For completeness and for future reference, we display below the 
complete expressions for the $\mathcal{J}^{\mu_1 \cdots \mu_\ell}$ terms
\cite{Denicol.2012},
\begin{widetext}
\begin{subequations} \label{eq:DNMR_J}
\begin{align}
 \mathcal{J} =& -\ell_{\Pi n} \nabla \cdot n - 
 \tau_{\Pi n} n \cdot F - \delta_{\Pi\Pi} \Pi \theta 
 - \lambda_{\Pi n} n \cdot I + 
 \lambda_{\Pi \pi} \pi^{\mu\nu} \sigma_{\mu\nu}\;,\\
 \mathcal{J}^\mu =& -\tau_n n_\nu \omega^{\nu\mu} - \delta_{nn} n^\mu \theta 
 - \ell_{n\Pi} \nabla^\mu \Pi + \ell_{n\pi} \Delta^{\mu\nu} \nabla_\lambda \pi^\lambda{}_\nu + \tau_{n\Pi} \Pi F^\mu - 
 \tau_{n\pi} \pi^{\mu\nu} F_\nu \nonumber\\
 & -\lambda_{nn} n_\nu \sigma^{\mu\nu} + 
 \lambda_{n \Pi} \Pi I^\mu - \lambda_{n\pi} \pi^{\mu\nu} I_\nu\;,\\
 \mathcal{J}^{\mu\nu} =& 2\tau_\pi\pi^{\langle\mu}_\lambda \omega^{\nu\rangle \lambda} - 
 \delta_{\pi\pi} \pi^{\mu\nu} \theta - 
 \tau_{\pi\pi} \pi^{\lambda\langle \mu}\sigma^{\nu\rangle}_\lambda + 
 \lambda_{\pi \Pi} \Pi \sigma^{\mu\nu} - 
 \tau_{\pi n} n^{\langle \mu} F^{\nu \rangle} + \ell_{\pi n} \nabla^{\langle \mu} n^{\nu \rangle} 
 + \lambda_{\pi n} n^{\langle \mu} I^{\nu \rangle}\;,
\end{align}
\end{subequations}
where $F^\mu = \nabla^\mu P$ and $I^\mu = \nabla^\mu \alpha$.
We also display the $\mathcal{K}^{\mu_1 \cdots \mu_\ell}$ terms, 
following the conventions of Ref.~\cite{Molnar.2014}:
\begin{subequations} \label{eq:DNMR_K}
\begin{align}
 \mathcal{K} =& \tilde{\zeta}_1   \omega_{\mu\nu} \omega^{\mu\nu} + 
 \tilde{\zeta}_2 \sigma_{\mu\nu} \sigma^{\mu\nu} + \tilde{\zeta}_3 \theta^2 + 
 \tilde{\zeta}_4 I \cdot I + \tilde{\zeta}_5 F \cdot F + 
\tilde{\zeta}_6 I \cdot F + \tilde{\zeta}_7 \nabla \cdot I + 
 \tilde{\zeta}_8 \nabla \cdot F\;,\\
 \mathcal{K}^\mu =& \tilde{\kappa}_1 \sigma^{\mu\nu} I_\nu + \tilde{\kappa}_2 \sigma^{\mu\nu} F_\nu + \tilde{\kappa}_3 I^\mu \theta + 
 \tilde{\kappa}_4 F^\mu \theta + \tilde{\kappa}_5 \omega^{\mu\nu} I_\nu + 
 \tilde{\kappa}_6 \Delta^\mu_\lambda \nabla_\nu \sigma^{\lambda \nu} + 
 \tilde{\kappa}_7 \nabla^\mu \theta\;,\\
 \mathcal{K}^{\mu\nu} =& \tilde{\eta}_1 \omega^{\lambda\langle \mu} \omega^{\nu\rangle}{}_\lambda + \tilde{\eta}_2 \theta \sigma^{\mu\nu} + 
 \tilde{\eta}_3 \sigma^{\lambda \langle \mu} \sigma^{\nu\rangle}_\lambda + 
 \tilde{\eta}_4 \sigma^{\langle \mu}_\lambda \omega^{\nu\rangle \lambda} + 
 \tilde{\eta}_5 I^{\langle\mu} I^{\nu \rangle} + \tilde{\eta}_6 F^{\langle\mu}
 F^{\nu\rangle} \nonumber\\
 & + \tilde{\eta}_7 I^{\langle\mu}F^{\nu\rangle} + 
 \tilde{\eta}_8 \nabla^{\langle\mu} I^{\nu \rangle} + 
 \tilde{\eta}_9 \nabla^{\langle \mu} F^{\nu \rangle}\;.
\end{align}
\end{subequations}
\end{widetext}

To understand the origin of the $\mathcal{O}(\Re \Kn)$ and $\mathcal{O}(\Kn^2)$ terms, we note that the asymptotic matching in Eqs.~\eqref{eq:DNMR_matching} replaces the irreducible moments [originally of order $\mathcal{O}(\Re)$] with $\mathcal{O}(\Re)$ and $\mathcal{O}(\Kn)$ terms proportional to $(\Pi, n^\mu, \pi^{\mu\nu})$ and $(\theta,I^\mu, \sigma^{\mu\nu})$, respectively. At the level of Eqs.~\eqref{eq:DNMR_eqs},
the former terms make $\mathcal{O}(\Re \Kn)$ contributions, while the latter ones give rise to $\mathcal{O}(\Kn^2)$ terms. This can be easily seen in what concerns the terms appearing on the right-hand side of Eqs.~\eqref{eq:rhodot_all}, since there the irreducible moments always come with $\mathcal{O}(\Kn)$ coefficients. Additional contributions arise from the comoving derivative of the irreducible moments appearing on the left-hand side of Eqs.~\eqref{eq:rhodot_all}. We illustrate this by considering the particular example of the tensor moments $\dot{\rho}_r^{\langle\mu\nu\rangle}$. Taking the comoving derivative of Eq.~\eqref{eq:DNMR_matching_2} leads to
\begin{multline}
 \dot{\rho}_r^{\langle\mu\nu\rangle} = \Omega^{(2)}_{r0}\dot{\pi}^{\langle\mu\nu\rangle}+\dot{\Omega}_{r0}^{(2)}\pi^{\mu\nu} + 2 D[\eta(\mathcal{C}^{(2)}_r -\Omega_{r0}^{(2)})] \sigma^{\mu\nu}\\
 + 2\eta(\mathcal{C}_r^{(2)}-\Omega_{r0}^{(2)})\dot{\sigma}^{\langle\mu\nu\rangle} + \mathcal{O}(\Re \Kn^2)\;,\label{eq:rho_dot_explicit}
\end{multline}
where $\mathcal{C}_r^{(2)} = \eta_r / \eta$ was introduced in Eqs.~\eqref{eq:C_def}. The first term in Eq.~\eqref{eq:rho_dot_explicit} gives rise to the relaxation time $\tau_\pi$ via Eq.~\eqref{eq:DNMR_rtimes_2}. To leading order, the comoving derivative $Df = \dot{f}$ of a thermodynamic function $f \equiv f(\alpha,\beta)$ is of order $\mathcal{O}(\Kn)$, since
\begin{align}
 \dot{f} &= \frac{\partial f}{\partial \alpha} \dot{\alpha} + \frac{\partial f}{\partial \beta} \dot{\beta} \nonumber\\
 &= \left( \mathcal{H} \frac{\partial f}{\partial\alpha} + \overline{\mathcal{H}} \frac{\partial f}{\partial\beta} \right) \theta + \mathcal{O}(\Re \Kn),
 \label{eq:dotf}
\end{align}
where $\mathcal{H}$ and $\overline{\mathcal{H}}$ are defined in Eq.~\eqref{eq:H_def}, while $\dot{\alpha}$ and $\dot{\beta}$ are given in Eqs.~(\ref{eq:alphadot++}a,b). Thus, the second term of Eq. \eqref{eq:rho_dot_explicit} is of order $\mathcal{O}(\Re \Kn)$, contributing to $\mathcal{J}^{\mu\nu}$. In contrast, the third and fourth terms are of order $\mathcal{O}(\Kn^2)$, thus contributing to $\mathcal{K}^{\mu\nu}$.

As mentioned in the introduction, the $\mathcal{K}^{\mu_1\cdots \mu_\ell}$ terms are 
traditionally ignored in the literature, either because they vanish in the 14 moment limit, or because they lead to parabolic equations of motion \cite{Denicol:2012vq}.
In the following section, we rederive the evolution equations \eqref{eq:DNMR_eqs} 
such that $\mathcal{K}^{\mu_1 \cdots \mu_\ell}$ vanish identically by construction.

\section{Inverse-Reynolds-Dominance (IReD) approach}\label{sec:IReD}

\begin{table*}
\begin{tabular}{|lcccc||lcccc|}
\hline\hline
\multicolumn{5}{|c||}{[IReD]\quad 
Diffusion: Relaxation times \phantom{\quad[IReD]}} &
\multicolumn{5}{c|}{[DNMR] \qquad\quad Diffusion: Inverse eigenvalues \phantom{\qquad\quad[DNMR]}}\\
$N_1$ & $\tau_{n,0} [\lambda_{\rm mfp}]$  
& $\tau_{n,2} [\lambda_{\rm mfp}]$ 
& $\tau_{n,3} [\lambda_{\rm mfp}]$ 
& $\tau_{n,4} [\lambda_{\rm mfp}]$ 
& $N_1$ & $[\chi^{(1)}_0]^{-1} [\lambda_{\rm mfp}]$ 
& $[\chi^{(1)}_2]^{-1} [\lambda_{\rm mfp}]$ 
& $[\chi^{(1)}_3]^{-1} [\lambda_{\rm mfp}]$ 
& $[\chi^{(1)}_4]^{-1} [\lambda_{\rm mfp}]$ \\\hline
1 & $9/4$ & - & - & - &
1 & $9/4$ & - & - & - \\
2 & $2.076$ & $2.419$ & - & - &
2 & $2.59$ & $1.629$ & - & - \\
3 & $2.076$ & $2.435$ & $2.565$ & -&
3 & $2.575$ & $1.961$ & $1.413$ & -\\
4 & $2.079$ & $2.438$ & $2.568$ & $2.680$ &
4 & $2.573$ & $1.85$ & $1.597$ & $1.304$ \\\hline
$\infty$ & $2.084$ & $2.440$ & $2.570$ & $2.681$ &
$\infty$ & $2.572$ & $1.847$ & $1.586$ & $1.451$ \\\hline\hline
\multicolumn{5}{|c||}{[IReD]\qquad Shear: Relaxation times \phantom{\qquad[IReD]}} &
\multicolumn{5}{c|}{[DNMR]\qquad\qquad Shear: Inverse eigenvalues \phantom{\qquad\qquad[DNMR]}}\\
$N_2$ & $\tau_{\pi,0} [\lambda_{\rm mfp}]$  
& $\tau_{\pi,1} [\lambda_{\rm mfp}]$ 
& $\tau_{\pi,2} [\lambda_{\rm mfp}]$ 
& $\tau_{\pi,3} [\lambda_{\rm mfp}]$ &
$N_2$ & $[\chi^{(2)}_0]^{-1} [\lambda_{\rm mfp}]$ 
& $[\chi^{(2)}_1]^{-1} [\lambda_{\rm mfp}]$ 
& $[\chi^{(2)}_2]^{-1} [\lambda_{\rm mfp}]$ 
& $[\chi^{(2)}_3]^{-1} [\lambda_{\rm mfp}]$ \\\hline
0 & $5/3$ & - & - & - &
0 & $5/3$ & - & - & - \\
1 & $1.649$ & $1.785$ & - & - &
1 & $2$ & $1.364$ & - & - \\
2 & $1.654$ & $1.788$ & $1.902$ & -&
2 & $2$ & $1.646$ & $1.241$ & -\\
3 & $1.655$ & $1.789$ & $1.902$ & $2.001$ &
3 & $2$ & $1.650$ & $1.477$ & $1.176$ \\\hline
$\infty$ & $1.656$ & $1.789$ & $1.902$ & $2.001$ &
$\infty$ & $2$ & $1.650$ & $1.484$ & $1.386$ \\\hline\hline
\end{tabular}
\caption{(left) Relaxation times $\tau_{n;r}$ and $\tau_{\pi;r}$ 
corresponding to the vector and tensor moments $\rho_r^\mu$ and 
$\rho^{\mu\nu}_r$, respectively, obtained for various values of the truncation 
orders $N_1$ and $N_2 = N_1 - 1$. (right) Inverse eigenvalues 
$[\chi^{(1)}_r]^{-1}$ and $[\chi^{(2)}_r]^{-1}$ shown in descending order.
The relaxation times and inverse eigenvalues are expressed in units 
of the mean free path $\lambda_{\rm mfp} = 1/(n \sigma)$, where $n$ is the 
local particle-number density and $\sigma$ is the (constant) collision cross-section.
\label{tbl:rtimes}
}
\end{table*}

In this section, we discuss the derivation of the evolution equations \eqref{eq:DNMR_eqs} 
for the case when the terms of second order with respect to $\Kn$ vanish,
$\mathcal{K}^{\mu_1 \cdots \mu_\ell} = 0$. The derivation is identical to that 
presented in the previous section, up to Eqs.~\eqref{eq:A_def}. The main difference 
compared to the DNMR approach is at the level of the asymptotic matching. 
In this section, we bypass the diagonalization of the collision matrix via 
the matrix $\Omega^{(\ell)}_{rn}$. 
Multiplying Eqs.~\eqref{eq:rhodot_all} by $\tau^{(\ell)}_{nr}$ and summing over $r$, 
we arrive at \cite{Jan}
\begin{subequations} \label{eq:rhodot_projected}
\begin{align}
 \sum_{r=0,\neq 1,2}^{N_0} \tau_{nr}^{(0)} \dot{\rho}_r + \rho_n 
 =&\, \frac{3}{m^2} \zeta_n \theta + \mathcal{O}(\Kn\,\Re)\;,\\
 \sum_{r=0,\neq 1}^{N_1} \tau_{nr}^{(1)} \dot{\rho}^{\langle \mu \rangle}_r + 
 \rho^\mu_n =& \,\kappa_n I^\mu + \mathcal{O}(\Kn\,\Re)\;,\\
 \sum_{r=0}^{N_2} \tau_{nr}^{(2)} \dot{\rho}^{\langle \mu\nu \rangle}_r + 
 \rho^{\mu\nu}_n =&\, 2\eta_n \sigma^{\mu\nu} + \mathcal{O}(\Kn\,\Re)\;,
\end{align}
\end{subequations}
where the first-order transport coefficients $\zeta_n$, $\kappa_n$ and $\eta_n$
were introduced in Eqs.~\eqref{eq:NS_coeff}.
Note that the comoving derivatives on the left-hand sides of
Eqs.~\eqref{eq:rhodot_projected} are of order $\mathcal{O}(\Kn\,\Re)$ as well.
Neglecting terms of this order, we obtain straightforwardly from Eqs.~\eqref{eq:rhodot_projected}
\begin{equation}
 \rho_n \simeq\, \frac{3}{m^2} \zeta_n \theta\;, \quad
 \rho^\mu_n \simeq\, \kappa_n I^\mu\;,\quad
 \rho^{\mu\nu}_n \simeq\, 2\eta_n \sigma^{\mu\nu}\;,
 \label{eq:C_matching_aux}
\end{equation}
while $\rho^{\mu\nu \lambda \cdots}_n \simeq \mathcal{O}(\Kn^2, \Kn\ \Re)$.
The above relations establish the correspondence between quantities of orders $\mathcal{O}(\Re)$ and $\mathcal{O}(\Kn)$ appearing on the left- and right-hand sides, respectively. We now exploit this correspondence in order to eliminate the $\mathcal{O}(\Kn)$ terms appearing in the DNMR matching prescription shown in Eqs.~\eqref{eq:DNMR_matching}.
Specializing the above relations to the case $n = 0$  and using Eqs.~\eqref{eq:zeroth moments} allows the 
thermodynamic forces $\theta$, $n^\mu$ and $\sigma^{\mu\nu}$ to be 
replaced by the dissipative quantities $\Pi$, $n^\mu$ and $\pi^{\mu\nu}$,
leading to the asymptotic matching equations
\begin{equation}
 \rho_n \simeq\, -\frac{3}{m^2} \mathcal{C}^{(0)}_n \Pi\;, \quad
 \rho^\mu_n \simeq\, \mathcal{C}^{(1)}_n n^\mu\;,\quad
 \rho^{\mu\nu}_n \simeq\, \mathcal{C}^{(2)}_n \pi^{\mu\nu}\;,
 \label{eq:C_matching}
\end{equation}
where the coefficients $\mathcal{C}^{(\ell)}_n$ were introduced in 
Eqs.~\eqref{eq:C_def}. 
Eqs.~\eqref{eq:C_matching} naturally hold also when 
$n = -r < 0$ by identifying
\begin{equation}\label{eq:negative_r_C}
 \mathcal{C}^{(\ell)}_{-r} = \bar{\gamma}^{(\ell)}_r = 
 \sum_{n = 0}^{N_\ell} \mathcal{F}^{(\ell)}_{rn} \mathcal{C}^{(\ell)}_n\;,
\end{equation}
where $\bar{\gamma}^{(\ell)}_r$ was introduced in Eqs.~\eqref{eq:DNMR_gamma_bar} and 
the function $\mathcal{F}^{(\ell)}_{rn}$ is defined in Eq.~\eqref{eq:DNMR_F}.
Equations \eqref{eq:C_matching} relate the higher-order moments 
$\rho_{r>0}^{\mu_1 \cdots\mu_\ell}$ to the zeroth-order ones. As mentioned in the introduction, a similar approach
was proposed under the name of the {\it order of magnitude} approach in Ref.~\cite{Struchtrup.2004} in the case of non-relativistic 
fluids, as well as in Ref.~\cite{Jan} for multicomponent relativistic fluids.
In the following, we will refer to this approach as the \emph{Inverse-Reynolds-Dominance} (IReD) approach, for reasons that will become apparent. 

We first remark that Eqs.~\eqref{eq:C_matching} is equivalent to the original 
DNMR matching in Eqs.~\eqref{eq:DNMR_matching}. This can be seen by replacing 
$\theta = -\Pi / \zeta$, $I^\mu = n^\mu / \kappa$ and $\sigma^{\mu\nu} = \pi^{\mu\nu} / (2\eta)$ and noting that the error introduced by these
replacements can be neglected since it is of higher order 
than the terms shown in Eqs.~\eqref{eq:DNMR_matching}. By using the relations \eqref{eq:C_matching} in the equations of motion \eqref{eq:rhodot_projected}, we can replace all irreducible moments appearing on the right-hand side by the 
%hydrodynamic variables 
dissipative quantities
$\Pi,\,n^\mu$, and $\pi^{\mu\nu}$, with the neglected terms being of order $\mathcal{O}(\Kn^2\,\Re)$. Furthermore, setting the index $n=0$ in Eqs.\ \eqref{eq:rhodot_projected}, we obtain the relaxation equations \eqref{eq:DNMR_eqs} with $\mathcal{K}^{\mu_1 \cdots \mu_\ell} = 0$. 
The $\mathcal{J}^{\mu_1\cdots \mu_\ell}$ terms retain the form in 
Eqs.~\eqref{eq:DNMR_J} and the transport coefficients arising there 
are identical in form to those derived in the DNMR formalism and reported in 
Ref.~\cite{Denicol.2012}, with the exception that all instances of $\Omega^{\ell}_{r0}$ 
should be replaced by $\mathcal{C}^{(\ell)}_r$ (also $\gamma^{(\ell)}_r$ should be 
replaced by $\bar{\gamma}^{(\ell)}_r \equiv \mathcal{C}^{(\ell)}_{-r}$):\footnote{See Appendix C of Ref.~\cite{Jan} for
explicit expressions in the case of a multicomponent fluid.}
\begin{subequations} \label{eq:connection}
\begin{align}
 &\text{(DNMR)} & & & 
 &\text{(IReD)} \nonumber\\
 &\Omega^{(\ell)}_{r0} & \boldsymbol{\longrightarrow}& & &\mathcal{C}^{(\ell)}_r\;,\\
 &\gamma^{(\ell)}_r & \boldsymbol{\longrightarrow}& & & \mathcal{C}^{(\ell)}_{-r}\;,\\
 &\mathcal{K}^{\mu_1 \cdots \mu_\ell} & \boldsymbol{\longrightarrow}& & & 0\;.
\end{align}
\end{subequations}
The expressions for the transport coefficients obtained using the IReD approach are summarized in Appendix~\ref{app:tcoeffs}. The above prescription holds also for the computation of the relaxation times. 
Replacing $\Omega^{(\ell)}_{r0}$ with
$\mathcal{C}^{(\ell)}_r$ in Eqs.~\eqref{eq:DNMR_rtimes}, we arrive at
\begin{subequations}\label{eq:C_rel_times}
\begin{align}
 \tau_\Pi =& \sum_{r = 0,\neq 1,2}^{N_0} \tau^{(0)}_{0r} \mathcal{C}^{(0)}_r\;,\\
 \tau_n =& \sum_{r = 0,\neq 1}^{N_1} \tau^{(1)}_{0r} \mathcal{C}^{(1)}_r\;,\\
 \tau_\pi =& \sum_{r = 0}^{N_2} \tau^{(2)}_{0r} \mathcal{C}^{(2)}_r\;.
 \label{eq:C_rel_times_2}
\end{align}
\end{subequations}
Upon performing the replacements in Eqs.~\eqref{eq:connection}, the values of the transport coefficients arising in the IReD approach will be different from those computed using the DNMR approach. This is clearly the case for the coefficients of the $\mathcal{O}(\Kn^2)$ terms, which vanish identically in the IReD approach. We will come back to the relation between the IReD and DNMR transport coefficients in the next section.

The matching procedure in Eqs.~\eqref{eq:C_matching}
eliminates the $\mathcal{K}^{\mu_1 \cdots \mu_\ell}$ terms which are 
of order $\mathcal{O}(\Kn^2)$, retaining the $\mathcal{J}^{\mu_1 \cdots \mu_\ell}$
terms of order $\mathcal{O}(\Kn\ \Re)$ and thereby trading one power of 
$\Kn$ for a power of $\Re$. This is clear when considering the terms appearing on the right-hand side of Eqs.~\eqref{eq:rhodot_all} [see also the discussion before Eq.~\eqref{eq:rho_dot_explicit}]. The comoving derivatives of the irreducible moments appearing on the left-hand side of Eqs.~\eqref{eq:rhodot_all} make only $\mathcal{O}(\Re \Kn)$ contributions. To see this, we reconsider the comoving derivative of the tensor moments with the asymptotic matching in Eqs.~\eqref{eq:C_matching},
\begin{equation}
    \dot{\rho}_r^{\langle\mu\nu\rangle} = \mathcal{C}_r^{(2)} \dot{\pi}^{\langle\mu\nu\rangle} + \dot{\mathcal{C}}_{r}^{(2)} \pi^{\mu\nu} + \mathcal{O}(\Re \Kn^2)\;.
    \label{eq:rho_dot_explicit_IReD}
\end{equation}
The first term contributes to the relaxation time $\tau_\pi$ via Eq.~\eqref{eq:C_rel_times_2}. As indicated in Eq.~\eqref{eq:dotf},  $\dot{\mathcal{C}}_r^{(2)}$ is of order $\mathcal{O}(\Kn)$, such that the second term is of order $\mathcal{O}(\Kn\ \Re)$, contributing only to $\mathcal{J}^{\mu\nu}$. 
We have thus established that the $\mathcal{O}(\Kn^2)$ terms vanish identically under the asymptotic matching in Eqs.~\eqref{eq:C_matching}. For this reason, we refer to this approach as the {\it Inverse-Reynolds-Dominance} (IReD) approach.

The connection between the IReD relaxation times in Eqs.~\eqref{eq:C_rel_times} and
the eigenvalues of $\mathcal{A}^{(\ell)}_{rn}$ is lost, therefore
one may wonder about the fate of the \textit{separation of scales}. 
In order to analyse the timescales associated with higher-order moments, it is 
convenient to introduce the coefficients $\mathcal{C}^{(\ell)}_{n;r}$ via
\begin{align}
 \mathcal{C}^{(0)}_{n;r} =& \frac{\zeta_n}{\zeta_r}\;, &
 \mathcal{C}^{(1)}_{n;r} =& \frac{\kappa_n}{\kappa_r}\;, &
 \mathcal{C}^{(2)}_{n;r} =& \frac{\eta_n}{\eta_r}\;,
\end{align}
such that $\mathcal{C}^{(\ell)}_{n;0} = \mathcal{C}^{(\ell)}_{n}$ reduces to 
the coefficients introduced in Eqs.~\eqref{eq:C_def}. To obtain the evolution 
equations for the irreducible moments $\rho^{\mu_1 \cdots \mu_\ell}_{r \neq 0}$, 
all the other irreducible moments should be written in terms of these ones 
via formulas analogous to Eqs.~\eqref{eq:C_matching},
\begin{align}
 \rho_n \simeq&\, \mathcal{C}^{(0)}_{n;r} \rho_r\;, &
 \rho^\mu_n \simeq&\, \mathcal{C}^{(1)}_{n;r} \rho^\mu_r\;, &
 \rho^{\mu\nu}_n \simeq&\, \mathcal{C}^{(2)}_{n;r} \rho^{\mu\nu}_r\;.
 \label{eq:C_matching_gen}
\end{align}
With these relations, we can apply the same procedure that was employed to yield 
Eqs.~\eqref{eq:DNMR_eqs} and obtain
\begin{subequations}\label{higher_order_rel}
\begin{align}
 \tau_{\Pi; r} \dot{\rho}_r + \rho_r =& \,\frac{3}{m^2} \zeta_r \theta + \mathcal{O}(\Kn \,\Re)\;,\\
 \tau_{n; r} \dot{\rho}^{\langle \mu\rangle}_r + \rho^\mu_r 
 =&\, \kappa_r I^\mu + \mathcal{O}(\Kn \,\Re)\;,\\
 \tau_{\pi; r} \dot{\rho}^{\langle \mu\nu \rangle}_r + \rho^{\mu\nu}_r 
 =&\, 2\eta_r \sigma^{\mu\nu} + \mathcal{O}(\Kn \,\Re)\;,
\end{align}
\end{subequations}
where the omitted terms on the right-hand side are of the same structure 
as Eqs.~\eqref{eq:DNMR_J}. The relaxation times appearing above are given 
by equations analogous to Eqs.~\eqref{eq:C_rel_times}, with 
$\mathcal{C}^{(\ell)}_{r} \equiv \mathcal{C}^{(\ell)}_{r;0}$ replaced by 
$\mathcal{C}^{(\ell)}_{r;n}$:
\begin{subequations}\label{eq:C_rel_times_gen}
\begin{align}
 \tau_{\Pi;n} =& \sum_{r=0,\neq 1,2}^{N_0} \tau^{(0)}_{nr} \mathcal{C}^{(0)}_{r;n}\;,\\
 \tau_{n;n} =& \sum_{r=0,\neq 1}^{N_1} \tau^{(1)}_{nr} \mathcal{C}^{(1)}_{r;n}\;,\\
 \tau_{\pi;n} =& \sum_{r=0}^{N_2} \tau^{(2)}_{nr} \mathcal{C}^{(2)}_{r;n}\;.
\end{align}
\end{subequations}
Setting $n=0$ in the above equations reproduces Eqs.\ \eqref{eq:C_rel_times}. 
The ordering of the relaxation times thus obtained clearly depends on the details 
of the (inverse of the) collision matrix. For definiteness, we report in 
Table~\ref{tbl:rtimes} the first four relaxation times in comparison 
to the first four eigenvalues $\chi^{(\ell)}_n$ obtained 
for the case of an ultrarelativistic ideal gas interacting via a constant cross-section
$\sigma$ (to be discussed in Sec.~\ref{sec:UR}). 
It can be seen that the {\it separation of scales} 
principle invoked in the
DNMR approach no longer holds, being
in fact reversed. The relaxation times obey the inequality 
$\tau_{*;0} \le \tau_{*; 1} \le \cdots$, for all $* \in \{n, \pi\}$ (the bulk sector does not contribute to the dynamics for a gas of massless particles). 

Based on the above analysis, it becomes evident that
demanding that the $\mathcal{O}(\Kn^2)$ terms vanish
gives relaxation times which are not compatible with the \textit{separation
of scales} concept. Conversely, enforcing
the {\it separation of scales} as done in DNMR (by setting $\tau_\Pi = [\chi_0^{(0)}]^{-1}$, etc) introduces in principle terms of order $\mathcal{O}(\Kn^2)$ 
in the evolution equations for the dissipative quantities. 
Despite this difference, the DNMR and the IReD approaches are equivalent, as we will show in the next section.

\begin{table}
\begin{tabular}{l|l}
IReD & DNMR \\\hline
%%% Relaxation times
\rule{0pt}{4ex}    
$\tau_\Pi$ & ${\displaystyle\tilde{\tau}_\Pi + \frac{\tilde{\zeta}_1}{\zeta}}$ \\[6pt]
\rule{0pt}{4ex}    
$\tau_n$ & ${\displaystyle \tilde{\tau}_n + \frac{\tilde{\kappa}_5}{2\kappa}}$\\[6pt]
\rule{0pt}{4ex}    
$\tau_\pi$ & ${\displaystyle \tilde{\tau}_\pi + \frac{\tilde{\eta}_1}{2\eta}}$ \\[9pt] \hline 
%%% Scalar coefficients
\rule{0pt}{4ex} 
$\ell_{\Pi n}$ & ${\displaystyle \tilde{\ell}_{\Pi n} - \frac{\tilde{\zeta}_7}{\kappa}}$\\
\rule{0pt}{4ex}    
$\tau_{\Pi n}$ & ${\displaystyle \tilde{\tau}_{\Pi n} 
-\frac{\tilde{\zeta}_1 D_{20}\mathcal{H}}{\kappa (\varepsilon+P)^3}
-\frac{\tilde{\zeta}_6}{\kappa}
-\frac{\tilde{\zeta}_7}{\kappa^2(\varepsilon+P)}\frac{\partial\kappa}{\partial\ln\beta}}$\\
\rule{0pt}{4ex}    
$\delta_{\Pi \Pi}$ & ${\displaystyle \tilde{\delta}_{\Pi \Pi} -\frac{\tilde{\zeta}_1}{\zeta^2}\left(
\mathcal{H}\frac{\partial\zeta}{\partial\alpha}+
%\frac{\dot{\beta}}{\theta}
\overline{\mathcal{H}}
\frac{\partial\zeta}{\partial\beta} - \frac{\zeta}{3}\right) + 
\frac{\tilde{\zeta}_3}{\zeta}}$\\
\rule{0pt}{4ex}    
$\lambda_{\Pi n}$ & ${\displaystyle \tilde{\lambda}_{\Pi n} -\frac{\tilde{\zeta}_4}{\kappa}+\frac{\tilde{\zeta}_7}{\kappa^2}\left(\frac{\partial\kappa}{\partial\alpha}+\frac{1}{h}\frac{\partial\kappa}{\partial\beta}\right)}$\\
\rule{0pt}{4ex}    
$\lambda_{\Pi \pi}$ & ${\displaystyle \tilde{\lambda}_{\Pi \pi} +\frac{\tilde{\zeta}_1+\tilde{\zeta}_2}{2\eta}}$\\
\hline
%%% Vector coefficients 
\rule{0pt}{4ex}    
$\delta_{nn}+\frac{\zeta}{\kappa}\lambda_{n\Pi}$ & 
${\displaystyle \tilde{\delta}_{nn} 
+ \frac{\zeta}{\kappa}\tilde{\lambda}_{n\Pi}
- \frac{\tilde{\kappa}_3}{\kappa} + 
\frac{\mathcal{H} \tilde{\kappa}_5 + 2\tilde{\kappa}_7}{2\kappa \zeta} \left(\frac{\partial\zeta}{\partial\alpha}+\frac{1}{h}\frac{\partial\zeta}{\partial\beta}\right)}$ \\
& \qquad ${\displaystyle -\frac{\tilde{\kappa}_5}{2\kappa^2}
\left(
\overline{\mathcal{H}}\frac{\partial \kappa}{\partial\beta} + 
\frac{\kappa}{h} \frac{\partial \mathcal{H}}{\partial \beta} +
\frac{\partial (\kappa \mathcal{H})}{\partial\alpha} - \frac{\kappa}{3}
 \right)}$\\
\rule{0pt}{4ex}    
$\ell_{n \Pi}$ & ${\displaystyle \tilde{\ell}_{n \Pi} + \frac{\mathcal{H}\tilde{\kappa}_5 + 2\tilde{\kappa}_7}{2\zeta}}$\\
\rule{0pt}{4ex}    
$\ell_{n\pi}$ & ${\displaystyle \tilde{\ell}_{n \pi} + \frac{\tilde{\kappa}_6}{2\eta}}$ \\
\rule{0pt}{4ex}    
$\tau_{n\Pi}$ & ${\displaystyle \tilde{\tau}_{n\Pi}
- \frac{\tilde{\kappa}_4}{\zeta} - 
\frac{\mathcal{H} \tilde{\kappa}_5 + 2 \tilde{\kappa}_7}{2\zeta^2(\varepsilon + P)}
\frac{\partial \zeta}{\partial \ln \beta} + 
\frac{\tilde{\kappa}_5/2\zeta}{\varepsilon + P} \frac{\partial(\beta \mathcal{H})}{\partial \beta}}$\\
\rule{0pt}{4ex}    
$\tau_{n\pi}$ &${\displaystyle \tilde{\tau}_{n\pi} - \frac{\tilde{\kappa}_2}{2\eta}
-\frac{\tilde{\kappa}_6}{2\eta^2(\varepsilon+P)}
\frac{\partial\eta}{\partial\ln\beta}}$\\ 
\rule{0pt}{4ex}    
${\displaystyle \lambda_{nn}+ \frac{2\eta}{\kappa}\lambda_{n\pi}}$ & ${\displaystyle \tilde{\lambda}_{nn}+ \frac{2\eta}{\kappa}\tilde{\lambda}_{n\pi} -\frac{\tilde{\kappa}_1}{\kappa} + \frac{\tilde{\kappa}_5}{2\kappa}
+\frac{\tilde{\kappa}_6}{\eta\kappa}
\left(\frac{\partial\eta}{\partial\alpha} +\frac{1}{h}\frac{\partial\eta}{\partial\beta}\right)}$ \\
% ${\qquad \displaystyle-\tilde{\kappa}_6 \left(\frac{\partial}{\partial\alpha} +\frac{1}{h}\frac{\partial}{\partial\beta}\right)\frac{1}{2\eta}}$ \\
\hline
%%% Tensor coefficients
\rule{0pt}{4ex}    
${\displaystyle \delta_{\pi\pi}+\frac{\zeta}{2\eta}\lambda_{\pi\Pi}}$ & 
${\displaystyle \tilde{\delta}_{\pi\pi}+\frac{\zeta}{2\eta}\tilde{\lambda}_{\pi\Pi}
+ \frac{\tilde{\eta}_1}{3\eta} - \frac{\tilde{\eta}_2}{2\eta}
-\frac{\tilde{\eta}_1}{2\eta^2}
\left(\mathcal{H}
\frac{\partial \eta}{\partial\alpha} +
 \overline{\mathcal{H}}
\frac{\partial \eta}{\partial\beta}\right)}$\\
\rule{0pt}{4ex}    
$\tau_{\pi\pi}$ & ${\displaystyle \tilde{\tau}_{\pi\pi}+\frac{\tilde{\eta}_1-\tilde{\eta}_3}{2\eta}}$\\
%
% \rule{0pt}{4ex}    
% $\lambda_{\pi\Pi}$ & ${\displaystyle \lambda_{\pi\Pi} }$\\
% %
\rule{0pt}{4ex}    
$\tau_{\pi n}$ & ${\displaystyle \tilde{\tau}_{\pi n}
+ \frac{\tilde{\eta}_1 D_{20}\mathcal{H}}{\kappa (\varepsilon+P)^3} -
\frac{\tilde{\eta}_7}{\kappa} 
-\frac{\tilde{\eta}_8}{\kappa^2(\varepsilon+P)}
\frac{\partial\kappa}{\partial\ln\beta}}$ \\
% \qquad ${\displaystyle+\frac{\tilde{\eta}_8}{\varepsilon+P}
% \frac{\partial}{\partial\ln\beta}\frac{1}{\kappa}}$ \\
%
\rule{0pt}{4ex}    
$\ell_{\pi n}$ & ${\displaystyle \tilde{\ell}_{\pi n} + \frac{\tilde{\eta}_8}{\kappa}}$ \\
\rule{0pt}{4ex}    
$\lambda_{\pi n}$ & ${\displaystyle \tilde{\lambda}_{\pi n}+ \frac{\tilde{\eta}_5}{\kappa}-\frac{\tilde{\eta}_8}{\kappa^2} \left(\frac{\partial\kappa}{\partial\alpha}+
\frac{1}{h}\frac{\partial\kappa}{\partial\beta}\right)}$\\
\hline
\end{tabular}
\caption{
Comparison between the transport coefficients arising in 
the IReD approach (left column) and those arising in the DNMR approach.
The partial derivatives are taken by considering $\alpha = \beta \mu$ and 
$\beta$ as independent variables and $h=(\varepsilon+P)/n$ is the specific enthalpy.
The notation $\mathcal{H}$ and $\overline{\mathcal{H}}$ is 
introduced in Eq.~\eqref{eq:H_def}.
\label{tbl:comp}
}
\end{table}

\section{Connection to DNMR}\label{sec:connection}

As discussed in Sections~\ref{sec:DNMR} and \ref{sec:IReD}, the IReD approach yields 
relaxation equations for $\Pi$, $n^\mu$ and $\pi^{\mu\nu}$ for which 
$\mathcal{K}^{\mu_1 \cdots \mu_\ell} = 0$. Since the DNMR and IReD approaches are both 
exact to second order in $\Kn$ and $\Re$, they must 
coincide up to (and including) terms of second order. In order 
to distinguish between the transport coefficients arising in the two approaches, 
we will use a tilde $\widetilde{\phantom{\eta}}$ to 
denote transport coefficients computed in the DNMR approach.
Keeping in mind that 
the first-order transport coefficients $\zeta_n$, $\kappa_n$ and $\eta_n$ are exactly the 
same in the two approaches, being given by Eqs.~\eqref{eq:NS_coeff}, the goal of this 
section is to prove the following equivalence:
\begin{subequations}
\begin{align}
 \tau_\Pi \dot{\Pi} - \mathcal{J} =& \,
 \tilde{\tau}_\Pi \dot{\Pi} - \widetilde{\mathcal{J}} - \widetilde{\mathcal{K}}\;,\\
 \tau_n \dot{n}^{\langle \mu\rangle} - \mathcal{J}^\mu =& \,
 \tilde{\tau}_n \dot{n}^{\langle \mu \rangle} - \widetilde{\mathcal{J}}^\mu - \widetilde{\mathcal{K}}^\mu\;,\\
 \tau_\pi \dot{\pi}^{\langle \mu\nu\rangle} - \mathcal{J}^{\mu\nu} =& \,
 \tilde{\tau}_\pi \dot{\pi}^{\langle \mu\nu \rangle} - \widetilde{\mathcal{J}}^{\mu\nu} - \widetilde{\mathcal{K}}^{\mu\nu}\;,
\end{align}
\end{subequations}
where the $\mathcal{O}(\Kn^2)$ terms are absent on the left-hand side by virtue of the IReD asymptotic matching.

The detailed comparison will be carried out in Appendix~\ref{app:useful formulae}. Here we will put forth the key points and focus on the modification of the relaxation times. This modification arises due to terms in $\widetilde{\mathcal{K}}^{\mu_1\cdots \mu_\ell}$ that originate from $\dot{\rho}^{\langle \mu_1 \cdots \mu_\ell \rangle}_r$. Focusing on the DNMR asymptotic matching for the case of the tensor moments, we multiply Eq.~\eqref{eq:rho_dot_explicit} by $\tau^{(2)}_{0r}$ and sum with respect to $r$:
\begin{multline}
    \sum_{r=0}^{N_2}\tau_{0r}^{(2)}\dot{\rho}^{\langle\mu\nu\rangle}_r =
    \dot{\pi}^{\langle\mu\nu\rangle} \sum_{r = 0}^{N_2} \tau^{(2)}_{0r} \Omega^{(2)}_{r0} \\
    +2 \eta \dot{\sigma}^{\langle\mu\nu\rangle} \sum_{r=0}^{N_2}\tau_{0r}^{(2)}(\mathcal{C}^{(2)}_r-\Omega^{(2)}_{r0})+\cdots,
\end{multline}
where we omitted second-order terms proportional to $\pi^{\mu\nu}$ and $\sigma^{\mu\nu}$ that lead to contributions to $\widetilde{\mathcal{J}}^{\mu\nu}$ and $\widetilde{\mathcal{K}}^{\mu\nu}$. The summation with respect to $r$ can be performed in favor of the DNMR and IReD relaxation times $\tilde{\tau}_\pi$ and $\tau_\pi$, introduced in Eqs.~\eqref{eq:DNMR_rtimes_2} and \eqref{eq:C_rel_times_2}, respectively. Performing the same steps for the scalar and vector moments, we arrive at
\begin{align}
 \sum_{r = 0,\neq 1,2}^{N_0} \tau^{(0)}_{0r} \dot{\rho}_r =&\, -\frac{3}{m^2}[\tilde{\tau}_\Pi \dot{\Pi} - 
 \zeta ({\tau}_\Pi - \tilde{\tau}_\Pi)
 \dot{\theta} + \cdots]\;,\nonumber\\
 \sum_{r = 0,\neq 1}^{N_1} \tau^{(1)}_{0r} \dot{\rho}^{\langle\mu\rangle}_r =&\, \tilde{\tau}_n \dot{n}^{\langle\mu\rangle}+
 \kappa ({\tau}_n - \tilde{\tau}_n)
 \dot{I}^{\langle\mu\rangle} + \cdots\;,\nonumber\\
 \sum_{r = 0}^{N_2} \tau^{(2)}_{0r} \dot{\rho}^{\langle\mu\nu\rangle}_r =&\, \tilde{\tau}_\pi \dot{\pi}^{\langle\mu\nu\rangle}+
 2\eta ({\tau}_\pi - \tilde{\tau}_\pi)
 \dot{\sigma}^{\langle\mu\nu\rangle} + \cdots\;.
 \label{eq:connection_aux}
\end{align}
Employing now the first-order (Navier-Stokes) relations
\begin{subequations} \label{eq:NS_matching}
\begin{align}
 \zeta \theta =& -\Pi + \mathcal{O}(\Kn^2, \Kn\, \Re)\;,\\
 \kappa I^\mu =& n^\mu + \mathcal{O}(\Kn^2, \Kn\, \Re)\;,\\
 2\eta \sigma^{\mu\nu} =& \pi^{\mu\nu} + \mathcal{O}(\Kn^2, \Kn\, \Re)\;,
\end{align}
\end{subequations}
to eliminate the thermodynamic forces in favor of the corresponding fluxes,
it can be seen that the second terms in Eqs.~\eqref{eq:connection_aux} 
lead to the replacement of the DNMR relaxation times $(\tilde{\tau}_\Pi, \tilde{\tau}_n, \tilde{\tau}_\pi)$ 
by the IReD ones $(\tau_\Pi, \tau_n, \tau_\pi)$, e.g.
\begin{multline}
 \tilde{\tau}_\pi \dot{\pi}^{\langle\mu\nu \rangle} + 2\eta (\tau_\pi - \tilde{\tau}_\pi) \dot{\sigma}^{\langle \mu\nu \rangle} \\= 
 \tau_\pi \dot{\pi}^{\langle\mu\nu \rangle} - 
 (\tau_\pi - \tilde{\tau}_\pi) \pi^{\mu\nu} \frac{\dot{\eta}}{\eta} + \cdots,
\end{multline}
where the neglected terms are of third order.

The above discussion hints that the key to connecting the DNMR transport coefficients 
to the IReD ones is to look at the comoving derivatives of $\theta$, $I^\mu$ and $\sigma^{\mu\nu}$. The full expressions are derived in Appendix~\ref{app:useful formulae}.
%and are given explicitly in Eq.~\eqref{eq:app_dotted}. 
Here we just reproduce the terms that hold the key to establishing the connection between the DNMR and IReD
relaxation times, namely
\begin{subequations} \label{eq:comoving_forces}
\begin{align}
 \dot{\theta} =&\, \omega^{\mu \lambda} \omega_{\mu \lambda} + \cdots\;,\\
 \dot{I}^{\langle\mu\rangle} =& -\omega^{\mu\nu} I_\nu + \cdots\;,\\
 \dot{\sigma}^{\langle\mu\nu\rangle} =& -\omega^{\lambda\langle \mu} \omega^{\nu\rangle}{}_\lambda + \cdots\;.\label{eq:comoving_forces_2}
\end{align} 
\end{subequations}
The terms shown on the right-hand sides have no correspondent in the $\widetilde{\mathcal{J}}^{\mu_1 \cdots \mu_\ell}$ terms (except for the case of $\omega^{\mu\nu} I_\nu$, which can be related to $\omega^{\mu\nu} n_\nu / \kappa$), therefore the coefficients of these terms appearing in $\widetilde{\mathcal{K}}^{\mu_1 \cdots \mu_\ell}$
will modify the relaxation times appearing on the left-hand side 
of Eqs.~\eqref{eq:DNMR_eqs}.
Focusing on the tensor sector, one can use Eq.~\eqref{eq:comoving_forces_2} together with $\sigma^{\mu\nu} \simeq  \pi^{\mu\nu}/2\eta$ to establish 
\begin{equation}
 \tilde{\eta}_1 \omega^{\lambda\langle \mu} \omega^{\nu\rangle}{}_\lambda \simeq - \frac{\tilde{\eta}_1}{2\eta} \dot{\pi}^{\langle\mu\nu\rangle} + \frac{\tilde{\eta}_1 \dot{\eta}}{2\eta^2} \pi^{\mu\nu} + \cdots\;,
\end{equation}
where the dots indicate the $\mathcal{O}(\Kn^2)$ terms which were omitted in Eq.~\eqref{eq:comoving_forces_2}. The coefficient $\tilde{\eta}_1 / 2\eta$ of $-\dot{\pi}^{\langle\mu\nu\rangle}$ represents exactly the difference between the IReD and DNMR relaxation times. Performing the same steps for the scalar sector, we arrive at
\begin{subequations}\label{eq:conn_tau}
\begin{align}
 {\tau}_\Pi =& \tilde{\tau}_\Pi + \frac{\tilde{\zeta}_1}{\zeta}\;,\\
 {\tau}_\pi =& \tilde{\tau}_\pi + \frac{\tilde{\eta}_1}{2\eta}\;.
\end{align}
In the case of the vector relaxation time, the term $\tilde{\kappa}_5 \omega^{\mu\nu} I_\nu$ must simultaneously account for the change of the relaxation time on the left hand side (in the term $\tilde{\tau}_n \dot{n}^{\langle\mu\rangle}$), as well as in the 
first term appearing in $\widetilde{\mathcal{J}}^\mu$, namely $\tilde{\tau}_n \omega^{\mu\nu} n_\nu$. Since both terms have equal weights, they get 
one half of $\tilde{\kappa}_5 \omega^{\mu\nu} I_\nu$ each, such that
\begin{equation}
 {\tau}_n = \tilde{\tau}_n + \frac{\tilde{\kappa}_5}{2\kappa}\;.
\end{equation}
\end{subequations}
Likewise, the term $\tilde{\eta}_4\sigma_{\lambda}^{\ \langle\mu}\omega^{\nu\rangle\lambda}$ in $\widetilde{\mathcal{K}}^{\mu\nu}$ 
acts by 
changing $\tilde{\tau}_\pi$ in the term
$2\tilde{\tau}_\pi\pi_{\lambda}^{\ \langle\mu}\omega^{\nu\rangle\lambda}$ 
appearing
in $\mathcal{\widetilde{J}^{\mu\nu}}$. 
The resulting relaxation time is indeed equal to 
$\tau_\pi$ given in Eqs.~\eqref{eq:conn_tau} by virtue of the 
equality $\tilde{\eta}_4=2\tilde{\eta}_1$ established by 
Eq.~(I22) of Ref.~\cite{Molnar.2014}.
%(notice that, from eq. I22 of \cite{Molnar.2014}, %$\tilde{\eta}_4=2\tilde{\eta}_1$).

The relations in \eqref{eq:conn_tau} can be explicitly checked by noting that \cite{Molnar.2014}
\begin{subequations}
\begin{align}
 \tilde{\zeta}_1 =& \sum_{r = 0, \neq 1,2}^{N_0} \tau^{(0)}_{0r} (\zeta_r - \Omega^{(0)}_{r0} \zeta) = \zeta({\tau}_\Pi - \tilde{\tau}_\Pi)\;,\\
 \tilde{\kappa}_5 =& 2\sum_{r = 0, \neq 1}^{N_1} \tau^{(1)}_{0r} (\kappa_r - \Omega^{(1)}_{r0} \kappa) = 2\kappa({\tau}_n - \tilde{\tau}_n)\;,\\
 \tilde{\eta}_1 =& 2\sum_{r = 0}^{N_2} \tau^{(2)}_{0r} (\eta_r - \Omega^{(2)}_{r0} \eta) = 2\eta({\tau}_\pi - \tilde{\tau}_\pi)\;,
\end{align}
\end{subequations}
where the DNMR (with tilde) and IReD (without tilde) relaxation times arise by virtue of Eqs.~\eqref{eq:DNMR_rtimes} and \eqref{eq:C_rel_times}, respectively.

Table~\ref{tbl:comp} summarizes the connection between the transport coefficients appearing 
in the IReD and DNMR formulations. While in this section we focused the discussion
only on the relaxation times, the procedure to obtain the results reported in
Table~\ref{tbl:comp} is similar in spirit, involving straightforward but tedious algebra,
which is sketched in Appendix~\ref{app:useful formulae}.

\section{Connection to Denicol {\it et al.} \cite{Denicol:2012vq}}\label{sec:connection2}

In this section, we discuss the connection with
Ref.~\cite{Denicol:2012vq}, where 
the parabolic $\mathcal{K}^{\mu_1 \cdots \mu_\ell}$ are 
eliminated in the context of multiple dynamic moments. 
Without reviewing all the details of this work, we recall 
only the matching formulas given in Eq.~(20) of Ref.~\cite{Denicol:2012vq}, 
\begin{subequations}\label{eq:multi_matching}
\begin{align}
 \rho^\mu_r =& \lambda^{(1)}_{r0} n^\mu + 
 \lambda^{(1)}_{r2} \rho_2^\mu\;,\label{eq:multi_matching:vector}\\
 \rho^{\mu\nu}_r =& \lambda^{(2)}_{r0} \pi^{\mu\nu} + 
 \lambda^{(2)}_{r1} \rho_1^{\mu\nu}\;,
 \label{eq:multi_matching:tensor}
\end{align}
which address only the vector and tensor moments,
since the work is focused on massless constituents for 
which the scalar moments are irrelevant. 
Eqs.~\eqref{eq:multi_matching:vector} and 
\eqref{eq:multi_matching:tensor} can be 
supplemented naturally with an equivalent equation
for the scalar moments,
\begin{equation}
 \rho_r = -\frac{3}{m^2}\lambda^{(0)}_{r0} \Pi + 
 \lambda^{(0)}_{r3} \rho_3\;.
 \label{eq:multi_matching:scalar}
\end{equation}
\end{subequations}
The coefficients $\lambda^{(\ell)}_{rs}$ appearing 
above are given in Eq.~(21) of Ref.~\cite{Denicol:2012vq}
for $\ell = 1, 2$ as
\begin{subequations}
\begin{align}
 \lambda^{(1)}_{r0} =& \frac{\Omega^{(1)}_{20} \kappa_r - \Omega^{(1)}_{r0} \kappa_2}{\Omega^{(1)}_{20} \kappa_0 - \kappa_2}\;, & 
 \lambda^{(1)}_{r2} =& \frac{\Omega^{(1)}_{r0} \kappa_0 - \kappa_r}{\Omega^{(1)}_{20} \kappa_0 - \kappa_2}\;, \\
 \lambda^{(2)}_{r0} =& \frac{\Omega^{(2)}_{10} \eta_r - \Omega^{(2)}_{r0} \eta_1}{\Omega^{(2)}_{10} \eta_0 - \eta_1}\;, & 
 \lambda^{(2)}_{r1} =& \frac{\Omega^{(2)}_{r0} \eta_0 - \eta_r}{\Omega^{(2)}_{10} \eta_0 - \eta_1}\;.
\end{align}
In the case of the scalar moments, the relevant coefficients read
\begin{align}
 \lambda^{(0)}_{r0} =& \frac{\Omega^{(0)}_{30} \zeta_r - \Omega^{(0)}_{r0} \zeta_3}{\Omega^{(0)}_{30} \zeta_0 - \zeta_3}\;, & 
 \lambda^{(0)}_{r3} =& \frac{\Omega^{(0)}_{r0} \zeta_0 - \zeta_r}{\Omega^{(0)}_{30} \zeta_0 - \zeta_3}\;.
\end{align}
\end{subequations}
As shown in Ref.~\cite{Denicol:2012vq}, the above
matching prescription succeeds in reproducing 
$\mathcal{K} = \mathcal{K}^\mu = \mathcal{K}^{\mu\nu} = 0$,
which is identical to the desideratum of our IReD approach.
The connection with the current approach can be established
by 
downgrading the moments $\rho_3$, $\rho_2^\mu$ and $\rho_1^{\mu\nu}$ from being dynamical (i.e., separate degrees of freedom) by
using the matching formulas $\rho_3 = -(3/m^2) \mathcal{C}^{(0)}_3 \Pi$, $\rho_2^\mu = \mathcal{C}^{(1)}_2 n^\mu$ and $\rho_1^{\mu\nu} = \mathcal{C}^{(2)}_1 \pi^{\mu\nu}$ given in Eqs.~\eqref{eq:C_matching}. Noting that 
\begin{subequations}
\begin{align}
    \lambda^{(0)}_{r0} + \mathcal{C}^{(0)}_3 \lambda^{(0)}_{r3} =& \,\mathcal{C}^{(0)}_r\;,\\
    \lambda^{(1)}_{r0} + \mathcal{C}^{(1)}_2 \lambda^{(1)}_{r2} =& \,\mathcal{C}^{(1)}_r\;,\\
    \lambda^{(2)}_{r0} + \mathcal{C}^{(2)}_1 \lambda^{(2)}_{r1} =& \,\mathcal{C}^{(2)}_r\;,
\end{align}
\end{subequations}
it is clear that Eqs.~\eqref{eq:multi_matching} reduce to 
Eqs.~\eqref{eq:C_matching} for all values of $r$.

\section{Explicit values in the ultrarelativistic limit}
\label{sec:UR}

\begin{table*}
\begin{tabular}{lccccccc}
     Number of moments &
     $\kappa$ & $\tau_n [\lambda_{\rm mfp}]$ & $\delta_{nn}[\tau_n]$ 
     & $\lambda_{n n}[\tau_n]$ 
     & $\lambda_{n\pi}[\tau_n]$ 
     & $\ell_{n\pi}[\tau_n]$ 
     & $\tau_{n\pi}[\tau_n]$ \\
     \hline
     14 &
     $3/(16\sigma)$
     & $9/4$ 
     & $1$
     & $3/5$
     & $\beta/20$
     & $\beta/20$
     & $\beta / 80 P$ \\
     23 &
     $21/(128\sigma)$
     & $2.0759$ 
     & $1$
     & $0.85806$
     & $0.067742 \beta$
     & $0.030645 \beta$
     & $0.0076613 \beta/ P$\\
     32 &
     $0.16054/\sigma$
     & $2.0761$ 
     & $1$
     & $0.88847$
     & $0.069060 \beta$
     & $0.029064 \beta$
     & $0.0072661 \beta / P $\\
     41 &
     $0.15959/\sigma$
     & $2.0794$ 
     & $1$
     & $0.89501$
     & $0.069240\beta$
     & $0.028677 \beta $
     & $0.0071692 \beta / P $\\\hline
     $\infty$ &
     $0.158925/\sigma$
     & $2.0838$
     & $1$
     & $0.89862$
     & $0.069273\beta$
     & $0.028371\beta$
     & $0.0070927 \beta / P$\\\hline
     \hline
\end{tabular}
\caption{Transport coefficients for the diffusion current $n^{\mu}$ arising 
in the IReD approach for an ultrarelativistic 
ideal gas interacting via a constant cross-section $\sigma$ 
for various truncation orders. 
We use the convention $N_0 = N_1 + 1 = N_2 + 2$ and the total number of moments 
is $5N_2 + 3N_1+ N_0+9$.}
\label{tbl:diffusion}
\end{table*}

\begin{table*}
\begin{tabular}{lccccccc}
     Number of moments &
     $\eta$ & $\tau_\pi [\lambda_{\rm mfp}]$ & $\tau_{\pi\pi}[\tau_\pi]$ 
     & $\lambda_{\pi n}[\tau_\pi]$ 
     & $\delta_{\pi\pi}[\tau_\pi]$ 
     & $\ell_{\pi n}[\tau_\pi]$ 
     & $\tau_{\pi n}[\tau_\pi]$ \\
     \hline
     14 &
     $4/(3\sigma \beta)$
     & $5/3$ 
     & $10/7$
     & $0$
     & $4/3$
     & $0$
     & $0$ \\
     23 &
     $14/(11\sigma \beta)$
     & $1.6494$ 
     & $1.6850$
     & $0.23622 / \beta$
     & $4/3$
     & $-0.47244/\beta$
     & $-0.47244/(\beta P)$\\
     32 &
     $1.2685/(\sigma \beta)$
     & $1.6540$ 
     & $1.6936$
     & $0.21580/\beta$
     & $4/3$
     & $-0.54342/\beta$
     & $-0.54342/(\beta P)$\\
     41 &
     $1.2678/(\sigma \beta)$
     & $1.6552$ 
     & $1.6944$
     & $0.20890/\beta$
     & $4/3$
     & $-0.56014/\beta$
     & $-0.56014 / (\beta P)$\\\hline
     $\infty$ &
     $1.2676/(\sigma \beta)$
     & $1.6557 $
     & $1.6945$
     & $0.20503/\beta$
     & $4/3$
     & $-0.56960 /\beta$
     & $-0.56960 /(\beta P)$\\\hline
     \hline
\end{tabular}
\caption{Same as Table~\ref{tbl:diffusion} for the shear stress $\pi^{\mu\nu}$.}
\label{tbl:shear}
\end{table*}

We now explicitly evaluate the IReD transport coefficients reported in Appendix~\ref{app:tcoeffs} for an 
ultrarelativistic ideal fluid of hard spheres, interacting via a constant 
cross-section $\sigma$. The procedure for performing the calculations 
is identical to the one introduced in Ref.~\cite{Denicol.2012} and will 
therefore not be repeated here. 
Following Ref.~\cite{Denicol.2012}, we report the 
values of the coefficients obtained by employing 
$14$, $23$, $32$ and $41$ moments. In addition,
we report convergence ($\infty$) values for the transport coefficients,
which are obtained by employing high-precision 
arithmetics using \textsc{Mathematica} \cite{mathematica}
with $N_0 - 2 = N_1 -1 = N_2 = 100$. 
The values of the transport coefficients related to the diffusion current $n^\mu$ 
and shear stress $\pi^{\mu\nu}$ are reported in Tables~\ref{tbl:diffusion} and 
\ref{tbl:shear}, respectively. 
The tables showing these transport coefficients for $0 \le N_2 \le 100$, as well as the relaxation time and inverse eigenvalues listed in Table~\ref{tbl:rtimes}, can be accessed as supplementary material \cite{supp}.
Naturally, we do not report transport coefficients 
for the bulk viscous pressure $\Pi$, since, for massless particles, the bulk sector does not make any contribution.

\section{Conclusion}\label{sec:conc}

In this paper, we considered the connection between the transport coefficients 
arising in the standard DNMR and the IReD approach.
%and those corresponding to a hydrodynamic 
%theory when the terms of second order with respect to the Knudsen number $\Kn$
%are ignored. 
%Setting these terms to $0$ rigorously via the inverse Reynolds 
%dominance (IReD) approach, 
We show that the transport coefficients appearing in the
$\mathcal{J}^{\mu_1 \cdots \mu_\ell}$ terms [accounting for all 
$\mathcal{O}(\Kn\, \Re)$ contributions] receive modifications coming from 
the original $\mathcal{K}^{\mu_1 \cdots \mu_\ell}$ terms. Moreover, the 
relaxation times in the IReD approach differ from the DNMR ones, being given 
as a combination of the DNMR relaxation time and a second-order transport 
coefficient coming from the $\mathcal{K}^{\mu_1 \cdots \mu_\ell}$ terms.

In the process of absorbing the $\mathcal{K}^{\mu_1 \cdots \mu_\ell}$ terms, 
we obtained relaxation times which are no longer constrained to satisfy the 
\textit{separation of scales}. In particular, for the case of the ultrarelativistic
hard sphere ideal gas, we found that the relaxation times of the dissipative 
quantities $\Pi$, $n^\mu$ and $\pi^{\mu\nu}$ are smaller than those corresponding 
to higher-order moments. For the same system, we also reported accurate values for 
all transport coefficients (corresponding to the limits $N_0, N_1, N_2 \rightarrow \infty$) appearing in the vector and tensor sectors.

Due to their parabolic nature, the $\mathcal{K}^{\mu_1 \cdots \mu_\ell}$ terms 
which are quadratic in $\Kn$ may lead to violations of causality, as pointed out 
in Refs.~\cite{Denicol.2012,Denicol:2012vq}, and are therefore customarily omitted.
Our work provides the foundation for hydrodynamical theories which are free of 
such terms, while retaining second-order accuracy with respect to $\Kn$
and $\Re$. The absence of parabolic terms in the IReD approach
may help in deriving the entropy current from kinetic theory. Such an analysis was performed in the 14-moment approximation \cite{Groot.1980, Israel.1979,Hiscock.1983,El:2008yy}, where the parabolic terms are absent also in the DNMR approach. Extending the analysis beyond 14 moments (e.g., when $N_\ell \rightarrow \infty$) remains an open problem representing an interesting avenue for future research.

\begin{acknowledgments}
We thank D. Rischke, E. Moln\'ar, J. Fotakis, and P. Aasha for fruitful 
discussions. 
D.W.\ acknowledges support by the Studienstiftung des deutschen Volkes 
(German Academic Scholarship Foundation). 
A.P. acknowledges the kind hospitality of the Institute for Theoretical Physics, Goethe University, Frankfurt am Main (Germany), where this work was completed.
V.E.A. gratefully acknowledges the support of
the Alexander von Humboldt Foundation through a Research
Fellowship for postdoctoral researchers.
The authors acknowledge support by the Deutsche Forschungsgemeinschaft (DFG, German Research Foundation) through the CRC-TR 211 'Strong-interaction matter under extreme conditions'– project number 315477589 – TRR 211. D.W. acknowledges the support by the State of Hesse within the Research Cluster ELEMENTS (Project ID 500/10.006).
The authors gratefully acknowledge Dr. Flotte for hospitality and fruitful discussions.
\end{acknowledgments}

\appendix
\section{Equivalence between IReD and DNMR}\label{app:useful formulae}
In this appendix we report the calculations leading to Table \ref{tbl:comp}. We will manipulate the terms appearing in the
$\widetilde{\mathcal{K}}$, $\widetilde{\mathcal{K}}^\mu$ and $\widetilde{\mathcal{K}}^{\mu\nu}$
terms \eqref{eq:DNMR_K} with the purpose of absorbing them into the 
corresponding $\widetilde{\mathcal{J}}^{\mu_1 \cdots \mu_\ell}$ terms, thus
inferring the connection to the coefficients obtained in the 
IReD approach. We will employ the same notation as in Sec. \ref{sec:connection}, by which the DNMR quantities will be denoted with a tilde $\widetilde{\phantom{1}}$.
The main idea is to trade one power of $\Kn$
for one power of $\Re$. This is done using the 
Navier-Stokes asymptotic matching \eqref{eq:NS_matching} 
between the thermodynamic fluxes $\Pi$, $n^\mu$ and $\pi^{\mu\nu}$ 
and the thermodynamic forces $\theta$, $I^\mu$ and $\sigma^{\mu\nu}$.

As already mentioned in Sec.~\ref{sec:connection}, all terms 
appearing in $\widetilde{\mathcal{K}}^{\mu_1 \cdots \mu_\ell}$ can be 
related to those appearing in $\widetilde{\mathcal{J}}^{\mu_1 \cdots \mu_\ell}$,
with the exception of $\tilde{\zeta}_1 \omega_{\mu\nu} \omega^{\mu\nu}$ and 
$\tilde{\eta}_1 \omega_\lambda{}^{\langle\mu} \omega^{\nu\rangle \lambda}$ 
appearing in $\widetilde{\mathcal{K}}$ and $\widetilde{\mathcal{K}}^{\mu\nu}$, respectively.
We also include here the $\tilde{\kappa}_5 \omega^{\mu\nu} I_\nu$ term for reasons
that will become apparent. These terms can be related with the comoving derivatives 
of the thermodynamic forces, as suggested in Eqs.~\eqref{eq:comoving_forces}.
We start this section by deriving this latter equation.

We first recall Eqs.~(39)--(41) from Ref.~\cite{Denicol.2012},
\begin{subequations}\label{eq:alphadot++}
\begin{align}
    \dot{\alpha}=&\mathcal{H}\theta +\frac{J_{20}\Pi\theta-J_{30}\partial_\mu n^\mu}{D_{20}}- 
  \frac{J_{20}}{D_{20}} \pi^{\mu\nu} \sigma_{\mu\nu}\;,\\
    \dot{\beta}=& \overline{\mathcal{H}} \theta 
    +\frac{J_{10}\Pi\theta-J_{20}\partial_\mu n^\mu}{D_{20}} 
    - \frac{J_{10}}{D_{20}} \pi^{\mu\nu} \sigma_{\mu\nu}\;,\\
    \dot{u}^\mu =& \frac{F^\mu + \nabla^\mu \Pi - 
\Delta^\mu_{\ \alpha} \nabla_\beta \pi^{\alpha\beta} - 
\Pi \dot{u}^\mu - \pi^{\mu\nu} \dot{u}_\nu}
{\varepsilon + P}\;,
\end{align}
\end{subequations}
where $\mathcal{H}$ (introduced in Eq.~(I18) of Ref.~\cite{Molnar.2014}) and 
$\overline{\mathcal{H}}$ are defined as 
\begin{subequations}
\begin{align}
 \mathcal{H}=\frac{J_{20}(\varepsilon+P)-J_{30}n}{D_{20}}\;,\\
 \overline{\mathcal{H}} = \frac{J_{10}(\varepsilon+P)-J_{20}n}{D_{20}}\;,
 \label{eq:H_def}
\end{align}
\end{subequations}
while $J_{nq}$ and $D_{nq}$ are introduced above Eq.~\eqref{eq:I_def}.

The comoving derivative of $\theta = \partial_\mu u^\mu$ can be computed as follows:
\begin{equation}
    \dot{\theta} = \partial_\mu \dot{u}^\mu - 
    (\partial_\mu u_\lambda)(\partial^\lambda u^\mu).
\end{equation}
Noting that 
$\partial_\mu \dot{u}^\mu = \nabla_\mu \dot{u}^\mu - \dot{u}_\mu \dot{u}^\mu$ and 
$(\partial_\mu u_\lambda)(\partial^\lambda u^\mu) = 
(\nabla_\mu u_\lambda)(\nabla^\lambda u^\mu)$, we find
\begin{equation}
\dot{\theta} =-\dot{u}\cdot \dot{u} + \nabla_\alpha \dot{u}^\alpha - (\nabla_\alpha u^\rho) (\nabla_\rho u^\alpha).
\end{equation}
In the case of $I^\mu = \nabla^\mu \alpha$, the comoving derivative gives
\begin{equation}
 \dot{I}^\mu = \dot{\Delta}^\mu_\nu \partial^\nu \alpha + 
 \nabla^\mu \dot{\alpha} - 
 (\nabla^\mu u^\nu) (\partial_\nu \alpha).
\end{equation}
Projecting the above using $\Delta^\mu_\nu$ and 
noting that $\Delta^\mu_\nu \dot{\Delta}^\nu_\lambda = 
-\dot{u}^\mu u_\lambda$, we arrive at 
\begin{equation}
 \dot{I}^{\langle \mu \rangle} =-\dot{u}^\mu \dot{\alpha}+\nabla^{\mu}\dot{\alpha}-(\nabla^\mu u^\nu) I_\nu.
\end{equation}
Finally, the comoving derivative of $\sigma^{\mu\nu} =\Delta^{\mu\nu}_{\alpha\beta} \partial^\alpha u^\beta$ 
can be written as
\begin{equation}
 \dot{\sigma}^{\mu\nu} = \dot{\Delta}^{\mu\nu}_{\alpha\beta} \partial^\alpha u^\beta + 
 \nabla^{\langle \mu} \dot{u}^{\nu \rangle} - 
 \Delta^{\mu\nu}_{\alpha\beta} (\partial^\alpha u^\lambda)(\partial_\lambda u^\beta).
 \label{eq:sigmadot_aux}
\end{equation}
Using $\nabla^\alpha u^\rho = \sigma^{\alpha\rho} + \omega^{\alpha\rho} + \frac{1}{3} \theta\Delta^{\alpha\rho}$,
the last term can be expressed as
\begin{equation}
 \Delta^{\mu\nu}_{\alpha\beta} (\partial^\alpha u^\lambda)(\partial_\lambda u^\beta) = 
 \sigma^{\lambda\langle\mu}\sigma^{\nu\rangle}_{\ \lambda}+\omega^{\lambda\langle\mu}\omega^{\nu\rangle}_{\ \lambda}+\frac{2}{3}\sigma^{\mu\nu}\theta.
\end{equation}
Projecting Eq.~\eqref{eq:sigmadot_aux} using 
$\Delta^{\mu\nu}_{\alpha\beta}$ and using 
$\dot{\Delta}^{\langle \mu\nu \rangle}_{\alpha\beta} = \dot{\Delta}^{\mu\nu}_{\alpha\lambda} u^\lambda u_\beta + 
\dot{\Delta}^{\mu\nu}_{\beta\lambda} u^\lambda u_\alpha$, we arrive at
\begin{equation}
\dot{\sigma}^{\langle \mu\nu \rangle} = -\dot{u}^{\langle\mu}\dot{u}^{\nu\rangle}+\nabla^{\langle\mu}\dot{u}^{\nu\rangle}-\sigma^{\lambda\langle\mu}\sigma^{\nu\rangle}_{\ \lambda}-\omega^{\lambda\langle\mu}\omega^{\nu\rangle}_{\ \lambda}-\frac{2}{3}\sigma^{\mu\nu}\theta\;,
\label{eq:sigmadot}
\end{equation}
where we also used the property $\dot{\Delta}^{\mu\nu}_{\alpha\beta} \partial^\alpha u^\beta = \dot{\Delta}^{\mu\nu}_{\alpha\beta} \nabla^\alpha u^\beta - \dot{u}^{\langle \mu} \dot{u}^{\nu \rangle}$.

Using Eqs.~\eqref{eq:alphadot++} to leading order in $\Kn$ and $\Re$ leads to:
\begin{align}
 \dot{\theta} =& 
 \omega^{\mu\nu}\omega_{\mu\nu} -\sigma^{\mu\nu}\sigma_{\mu\nu} -\frac{1}{3}\theta^2 \nonumber\\
 &-\frac{2(\varepsilon + P) + \beta J_{30}}{(\varepsilon+P)^3}F\cdot F
 -\frac{D_{20} \mathcal{H}}{(\varepsilon+P)^3} I\cdot F\nonumber\\
 & + \frac{\nabla\cdot F}{\varepsilon+P}\;,\nonumber\\
 \dot{I}^{\langle\mu\rangle} =&
 -\sigma^{\mu\nu} I_\nu +I^\mu\theta\left(\frac{\partial\mathcal{H}}{\partial\alpha}+\frac{1}{h}\frac{\partial\mathcal{H}}{\partial{\beta}} - \frac{1}{3}\right) \nonumber\\
 & -\frac{F^\mu\theta}{\varepsilon + P} \frac{\partial(\beta \mathcal{H})}{\partial \beta}  
 - \omega^{\mu\nu} I_\nu
 +\mathcal{H}\nabla^\mu\theta\;,\nonumber\\
 \dot{\sigma}^{\langle\mu\nu\rangle} =& 
 -\omega^{\lambda\langle\mu}\omega^{\nu\rangle}{}_{\lambda}
 -\frac{2}{3}\theta\sigma^{\mu\nu}
 -\sigma^{\lambda\langle\mu}\sigma^{\nu\rangle}{}_{\lambda}
 \nonumber\\
 & -\frac{2(\varepsilon + P) + \beta J_{30}}
 {(\varepsilon+P)^3} F^{\langle\mu}F^{\nu\rangle}
 -\frac{D_{20} \mathcal{H}}{(\varepsilon+P)^3} I^{\langle\mu} F^{\nu\rangle} \nonumber\\
 & + \frac{\nabla^{\langle\mu} F^{\nu \rangle}}
 {\varepsilon + P}\;.\label{eq:app_dotted}
\end{align}

Using Eqs. (A10) to replace $\omega^{\mu\nu} \omega_{\mu\nu}$, $\omega^{\mu\alpha} I_\alpha$, and $\omega^{\lambda \langle \mu} \omega^{\nu \rangle}{}_{\lambda}$ in Eqs.~\eqref{eq:DNMR_K} gives
\begin{widetext}
\begin{subequations}
\begin{align}
 \widetilde{\mathcal{K}} =& \tilde{\zeta}_1   \dot{\theta} + 
 \left(\tilde{\zeta}_2+\tilde{\zeta}_1\right) \sigma_{\mu\nu} \sigma^{\mu\nu} + \left(\tilde{\zeta}_3+\frac{1}{3}\tilde{\zeta}_1\right) \theta^2 + 
 \tilde{\zeta}_4 I \cdot I + \left(\tilde{\zeta}_5+\tilde{\zeta}_1\frac{2(\varepsilon + P) + \beta J_{30}}{(\varepsilon+P)^3}\right) F \cdot F  \nonumber \\
 &+ \left(\tilde{\zeta}_6+\tilde{\zeta}_1
 %\frac{(\varepsilon + P) J_{20} - n J_{30}}{(\varepsilon+P)^3}
 \frac{D_{20}\mathcal{H}}{(\varepsilon+P)^3}\right) I \cdot F + \tilde{\zeta}_7 \nabla \cdot I + 
 \left(\tilde{\zeta}_8 -\frac{\tilde{\zeta}_1}{\varepsilon+P}\right)\nabla \cdot F\;,\\
\widetilde{\mathcal{K}}^\mu =&-\frac{\tilde{\kappa}_5}{2}\dot{I}^\mu 
+ \left(\tilde{\kappa}_1-\frac{\tilde{\kappa}_5}{2}\right) 
\sigma^{\mu\nu} I_\nu 
+ \tilde{\kappa}_2 \sigma^{\mu\nu} F_\nu 
+ \left[\tilde{\kappa}_3+\frac{\tilde{\kappa}_5}{2}\left(
\frac{\partial\mathcal{H}}{\partial\alpha}
+\frac{1}{h}\frac{\partial\mathcal{H}}{\partial{\beta}} - \frac{1}{3}\right)\right] I^\mu \theta  \nonumber\\
 &+ \left[\tilde{\kappa}_4 -\frac{\tilde{\kappa}_5}{2(\varepsilon+P)}
 \frac{\partial(\beta \mathcal{H})}{\partial \beta}\right] F^\mu \theta 
 +\frac{\tilde{\kappa}_5}{2}\omega^{\mu\nu} I_\nu 
 + \tilde{\kappa}_6 \Delta^\mu_\lambda \nabla_\nu \sigma^{\lambda \nu} 
 + \left(\tilde{\kappa}_7+\frac{\tilde{\kappa}_5}{2}\mathcal{H}\right) \nabla^\mu \theta\;,\\
 \widetilde{\mathcal{K}}^{\mu\nu} =& -\tilde{\eta}_1\dot{\sigma}^{\langle \mu\nu \rangle}  + \left(\tilde{\eta}_2-\frac{2}{3}\tilde{\eta}_1\right) \theta \sigma^{\mu\nu} + 
 \left(\tilde{\eta}_3-\tilde{\eta}_1\right) \sigma^{\lambda \langle \mu} \sigma^{\nu\rangle}_\lambda +  \tilde{\eta}_4 \sigma^{\langle \mu}_\lambda \omega^{\nu\rangle \lambda} + 
 \tilde{\eta}_5 I^{\langle\mu} I^{\nu \rangle}\nonumber\\
 &+ \left(\tilde{\eta}_6-\tilde{\eta}_1\frac{2(\varepsilon + P) + \beta J_{30}}{(\varepsilon+P)^3}\right) F^{\langle\mu}
 F^{\nu\rangle} 
 + \left(\tilde{\eta}_7-\tilde{\eta}_1
 %\frac{(\varepsilon + P) J_{20} - n J_{30}}{(\varepsilon+P)^3}
 \frac{D_{20}\mathcal{H}}{(\varepsilon+P)^3}
 \right) I^{\langle\mu}F^{\nu\rangle}\nonumber \\
 & + \tilde{\eta}_8 \nabla^{\langle\mu} I^{\nu \rangle} + 
 \left(\tilde{\eta}_9+\frac{\tilde{\eta}_1}{\varepsilon+P}\right) \nabla^{\langle \mu} F^{\nu \rangle}\;.
\end{align}
\end{subequations}
\end{widetext}
Using the relations (I5) and (I8) in Ref.~\cite{Molnar.2014}
relating $\tilde{\zeta}_5$ and $\tilde{\zeta}_8$ to 
$\tilde{\zeta}_1$, one can see that the coefficients in 
front of $F\cdot F$ and $\nabla\cdot F$ vanish identically.
Similarly, the relations (I24) and (I27) in Ref.~\cite{Molnar.2014} 
between $\tilde{\eta}_6$, $\tilde{\eta}_9$ and 
$\tilde{\eta}_1$ imply that the coefficients in front of 
$F^{\langle \mu}F^{\nu\rangle}$ and $\nabla^{\langle\mu}F^{\nu\rangle}$ 
also vanish. 
This is consistent with, and indeed required by, the 
equivalence between the IReD and DNMR approaches, since 
no such terms appear in either $\mathcal{J}$ or $\mathcal{J}^{\mu\nu}$. 
For this reason, the coefficients 
$\tilde{\zeta}_5$, $\tilde{\zeta}_8$, 
$\tilde{\eta}_6$ and $\tilde{\eta}_9$ do not 
appear in Table~\ref{tbl:comp}.

Comparing the above to Eqs.~\eqref{eq:DNMR_K}, it can be seen that 
aside from the new terms proportional to $\dot{\theta}$, $\dot{I}^{\langle \mu\rangle}$ 
and $\dot{\sigma}^{\langle \mu\nu \rangle}$, the coefficients of these terms 
($\tilde{\zeta}_1$, $\tilde{\kappa}_5$ and $\tilde{\eta}_1$) appear in several other 
terms. To compare with the coefficients obtained in the IReD approach, the thermodynamic forces $\theta$, $I^\mu$ and $\sigma^{\mu\nu}$ can be expressed in terms of the thermodynamic fluxes $\Pi$, $n^\mu$ and $\pi^{\mu\nu}$ via the asymptotic Navier-Stokes constitutive relations in Eqs.~\eqref{eq:NS_matching}. 
During this procedure, the comoving derivatives of the thermodynamic forces give rise to comoving derivatives of the thermodynamic fluxes, as well as to derivatives of the transport coefficients:
\begin{subequations}\label{eq:app_comoving}
\begin{align}
 \dot{\theta} =& -\frac{1}{\zeta} \dot{\Pi} + \frac{\Pi}{\zeta^2} \dot{\zeta}\;,\\
 \dot{I}^{\langle\mu\rangle} =& \frac{1}{\kappa} \dot{n}^{\langle \mu \rangle} 
 - \frac{n^\mu}{\kappa^2} \dot{\kappa}\;,\\
 \dot{\sigma}^{\langle \mu\nu\rangle} =& \frac{1}{2\eta} \dot{\pi}^{\langle\mu\nu\rangle}
 - \frac{\pi^{\mu\nu}}{2\eta^2} \dot{\eta}\;,
\end{align}
\end{subequations}
where the comoving derivative of a function depending on the fluid properties 
$\beta$ and $\alpha$ can be computed via Eq.~\eqref{eq:dotf}.
% \begin{equation}
%  \dot{f} \simeq 
%  \left( \overline{\mathcal{H}} \frac{\partial f}{\partial\beta}
%  +\mathcal{H} \frac{\partial f}{\partial\alpha}\right) \theta\;.
%  \end{equation}

The emergence of comoving derivatives of the thermodynamic forces in Eqs.~\eqref{eq:app_comoving} leads to modifications of the relaxation times $\tau_\Pi$, $\tau_n$ and $\tau_\pi$, as indiciated in Eqs.~\eqref{eq:conn_tau}. 
Furthermore, since the quantities in $\widetilde{\mathcal{K}}^{\mu_1\cdots\mu_\ell}$ are of second order in $\Kn$, the matching in Eqs.~\eqref{eq:NS_matching} reduces them to quantities of order $\mathcal{O}(\Kn\,\Re)$, which are then absorbed into the $\widetilde{\mathcal{J}}^{\mu_1 \cdots \mu_\ell}$ terms. 
By this procedure, the original transport coefficients appearing in $\widetilde{\mathcal{J}}^{\mu_1 \cdots \mu_\ell}$ are modified. Since the procedure stays accurate at second order with respect to $\Kn$ 
and $\Re$, the modified transport coefficients must exactly agree with those obtained in the IReD approach.
To illustrate the connection between the original and the modified transport coefficients,
let us focus on some examples concerning the terms in $\widetilde{\mathcal{K}}^\mu$. Starting from
\begin{equation}
    \tilde{\kappa}_6\Delta^{\mu}_{\ \lambda}\nabla_\nu\sigma^{\lambda\nu}\simeq 
    \frac{\tilde{\kappa}_6}{2\eta}\Delta^{\mu}{}_{\lambda}\nabla_\nu\pi^{\lambda\nu}
    -\tilde{\kappa}_6\frac{\pi^{\mu\nu}}{2\eta^2}\nabla_\nu\eta\;,
    \label{eq:app_k6}
\end{equation}
it can be seen that the first term is of the same form as
$\tilde{\ell}_{n\pi}\Delta^{\mu}_{\ \lambda}\nabla_\nu\pi^{\lambda\nu}$
and will thus lead to the following modification of this transport coefficient:
\begin{equation}
    {\ell}_{n\pi}=\tilde{\ell}_{n\pi}+\frac{\tilde{\kappa}_6}{2\eta}\;.
    \label{eq:app_lnpi}
\end{equation}
The relation above can be validated using the explicit expressions 
for $\tilde{\ell}_{n\pi}$ and $\tilde{\kappa}_6$, 
which we reproduce from Eq.~(C9) in Ref.~\cite{Denicol.2012} and 
Eq.~(I15) in Ref.~\cite{Molnar.2014}, respectively:
\begin{subequations}
\begin{align}
 \tilde{\ell}_{n\pi} =& - \tau_{00}^{(1)} \gamma_1^{(2)} +
 \sum_{r = 0,\neq 1}^{N_1} \tau^{(1)}_{0r} 
 \frac{\beta J_{r+2,1}}{\varepsilon + P} \nonumber\\
 &  - 
 \sum_{r = 0}^{N_1 - 2} \tau^{(1)}_{0,r+2} \Omega^{(2)}_{r+1,0}\;,\label{eq:lnpi}\\
 \tilde{\kappa}_6 =& -2 \sum_{r = 1}^{N_1 - 1} \tau^{(1)}_{0,r+1} 
 (\eta_{r} - \Omega^{(2)}_{r0} \eta)\;.\label{eq:k6}
\end{align}
\end{subequations}
Replacing $\gamma^{(2)}_1$ in the expression for $\tilde{\ell}_{n\pi}$ 
with $\bar{\gamma}^{(2)}_1$ defined in 
Eqs.~\eqref{eq:DNMR_gamma_bar} (see also the discussion around this equation),
we arrive at 
\begin{multline}
 \tilde{\ell}_{n\pi}+\frac{\tilde{\kappa}_6}{2\eta} = 
 - \tau_{00}^{(1)} \bar{\gamma}_1^{(2)} +
 \sum_{r = 0,\neq 1}^{N_1} \tau^{(1)}_{0r} 
 \frac{\beta J_{r+2,1}}{\varepsilon + P} \nonumber\\
 - \sum_{r = 0}^{N_1 - 2} \tau^{(1)}_{0,r+2} \mathcal{C}^{(2)}_{r+1},
\end{multline}
which is exactly the expression for ${\ell}_{n\pi}$ following 
the identification given in Eqs.~\eqref{eq:connection} applied to Eq.~\eqref{eq:lnpi} [see also Eq.~\eqref{eq:l_n_pi}].

The second term in Eq.~\eqref{eq:app_k6} gives rise to:
\begin{equation}
    \pi^{\mu\nu} \nabla_\nu\eta = 
    -\frac{\pi^{\mu\nu} F_\nu}{\varepsilon+P} 
    \frac{\partial\eta}{\partial\ln\beta} + 
    \pi^{\mu\nu} I_\nu
    \left(\frac{\partial\eta}{\partial\alpha}
    +\frac{1}{h}\frac{\partial\eta}{\partial\beta}\right).
\end{equation}
The terms on the right-hand side have the same form as the terms
$-\tilde{\lambda}_{n\pi} \pi^{\mu\nu} I_\nu$ and 
$-\tilde{\tau}_{n\pi} \pi^{\mu\nu} F_\nu$ appearing in 
$\mathcal{J}^\mu$, thus leading to a modification to these latter two 
transport coefficients ($\tilde{\lambda}_{n\pi}$ and $\tilde{\tau}_{n\pi}$).

It is worth noting that using the above procedure may lead to ambiguities. To 
illustrate such situations, let us focus on the term 
$\tilde{\kappa}_1 \sigma^{\mu\nu} I_\nu$, which can contribute to both
$-\tilde{\lambda}_{n\pi} \pi^{\mu\nu} I_\nu$ and to
$-\tilde{\lambda}_{nn} n_\nu \sigma^{\mu\nu}$, since
\begin{equation}
 \sigma^{\mu\nu}I_\nu \simeq \frac{\pi^{\mu\nu}}{2\eta} I_\nu \simeq \sigma^{\mu\nu}\frac{n_\nu}{\kappa}\;.
\end{equation}
Taking the first equality would modify only $\tilde{\lambda}_{n\pi}$, 
whereas taking the second equality modifies $\tilde{\lambda}_{nn}$. 
The decision on how to distribute the contribution from $\tilde{\kappa}_1$ to 
$\tilde{\lambda}_{n\pi}$ and $\tilde{\lambda}_{nn}$ can in principle be made 
by looking at the explicit expression for $\tilde{\kappa}_1$, reported in 
Eq.~(I10) of Ref.~\cite{Molnar.2014}. Another possibility is to acknowledge 
that this apparent ambiguity can be identified also in the form of 
$\widetilde{\mathcal{J}}^\mu$, allowing the two terms $\tilde{\lambda}_{n\pi} \pi^{\mu\nu} I_\nu$ 
and $\tilde{\lambda}_{nn} \sigma^{\mu\nu} n_\nu$ to be merged into a single one:
\begin{align}
\tilde{\lambda}_{nn} \sigma^{\mu\nu} n_\nu + \tilde{\lambda}_{n\pi} \pi^{\mu\nu}I_\nu &\simeq \left(\frac{\kappa}{2\eta} \tilde{\lambda}_{nn} + \tilde{\lambda}_{n\pi}\right) \pi^{\mu\nu} I_\nu  \nonumber\\
&\simeq \left(\tilde{\lambda}_{nn} + \frac{2\eta}{\kappa} \tilde{\lambda}_{n\pi}\right) 
\sigma^{\mu\nu}n_\nu\;.
\end{align}
Choosing to express all terms in the form $\sigma^{\mu\nu} n_\nu$, we obtain:
\begin{multline}
 {\lambda}_{nn}+\frac{2\eta}{\kappa}{\lambda}_{n\pi} = \tilde{\lambda}_{nn}+\frac{2\eta}{\kappa}\tilde{\lambda}_{n\pi}+ \frac{\tilde{\kappa}_6}{\eta \kappa}
 \left(\frac{\partial \eta}{\partial \alpha} + 
 \frac{1}{h} \frac{\partial \eta}{\partial \beta}\right)\\
 -\frac{1}{\kappa}\left(\tilde{\kappa}_1-\frac{\tilde{\kappa}_5}{2}\right).
 \label{eq:app_lambdann}
\end{multline}

The above discussion summarizes the key points required to obtain the 
relations presented in  Table~\ref{tbl:comp}. While Eqs.~\eqref{eq:app_lnpi}
and \eqref{eq:app_lambdann} refer only to the modifications of the  ${\ell}_{n\pi}$,
${\lambda}_{nn}$ and ${\lambda}_{n\pi}$ coefficients, the relations involving
the other coefficients can be derived following the same steps using straightforward
but lengthy algebra, which we do not present here explicitly.

\section{Second-order transport coefficients in the IReD approach}\label{app:tcoeffs}
In this appendix we give the transport coefficients of the IReD formalism. In what follows, we identify $\mathcal{C}_{-n}^{(\ell)}=\bar{\gamma}_n$, as in Eq. \eqref{eq:negative_r_C}. For the bulk pressure we have:
\begin{align}
    \ell_{\Pi n} =& -\frac{m^2}{3}\sum_{r=0,\neq 1,2}^{N_0}\tau_{0r}^{(0)}\left(\mathcal{C}^{(1)}_{r-1}-\frac{G_{3r}}{D_{20}}\right)\;,\\
    \tau_{\Pi n} =& \sum_{r=0,\neq 1,2}^{N_0} \frac{m^2 \tau_{0r}^{(0)}}{3(\varepsilon+P)} \left( r \mathcal{C}^{(1)}_{r-1}+ \frac{\partial\mathcal{C}_{r-1}^{(1)}}{\partial\ln \beta}-\frac{G_{3r}}{D_{20}}\right)\;,\\
    \delta_{\Pi\Pi} =&\sum_{r=0,\neq 1,2}^{N_0} \tau_{0r}^{(0)}\Bigg[ \frac{r+2}{3} \mathcal{C}^{(0)}_r + \mathcal{H}\frac{\partial\mathcal{C}^{(0)}_r}{\partial\alpha}+\overline{\mathcal{H}}\frac{\partial\mathcal{C}_r^{(0)}}{\partial\beta}\nonumber\\
    & -\frac{m^2}{3}(r-1)\mathcal{C}^{(0)}_{r-2}-\frac{m^2}{3} \frac{G_{2r}}{D_{20}} \Bigg]\;,\\
    \lambda_{\Pi n} = &-\frac{m^2}{3}\sum_{r=0,\neq 1,2}^{N_0}\tau_{0r}^{(0)}\left(\frac{\partial\mathcal{C}_{r-1}^{(1)}}{\partial\alpha}+\frac{1}{h}\frac{\partial\mathcal{C}_{r-1}^{(1)}}{\partial\beta}\right)\;,\\
    \lambda_{\Pi \pi} = &-\frac{m^2}{3}\sum_{r=0,\neq 1,2}^{N_0}\tau_{0r}^{(0)}\left[\frac{G_{2r}}{D_{20}}+(r-1)\mathcal{C}_{r-2}^{(2)}\right]\;. 
\end{align}

For the particle-diffusion current:
\begin{align}
    \delta_{nn} =& \sum_{r=0,\neq 1}^{N_1}\tau_{0r}^{(1)}\Bigg[
    \frac{r+3}{3} \mathcal{C}^{(1)}_r + 
    \mathcal{H}\frac{\partial\mathcal{C}_r^{(1)}}{\partial\alpha}+\bar{\mathcal{H}}\frac{\partial\mathcal{C}_r^{(1)}}{\partial\beta}\nonumber\\
    &-\frac{m^2}{3}(r-1)\mathcal{C}^{(1)}_{r-2}\Bigg]\;,\\
    \ell_{n\Pi}=&\sum_{r=0,\neq 1}^{N_1}\tau_{0r}^{(1)}\left(\frac{\beta J_{r+2,1}}{\varepsilon+P} - \mathcal{C}^{(0)}_{r-1}+\frac{1}{m^2} \mathcal{C}_{r+1}^{(0)}\right)\;,\\
    \ell_{n\pi}=&\sum_{r=0,\neq 1}^{N_1}\tau_{0r}^{(1)}\left(\frac{\beta J_{r+2,1}}{\varepsilon+P}-\mathcal{C}^{(2)}_{r-1}\right)\;,\label{eq:l_n_pi}\\
     \tau_{n\Pi}=&\sum_{r=0,\neq 1}^{N_1} \frac{\tau_{0r}^{(1)}}{\varepsilon+P} \left[\frac{\beta J_{r+2,1}}{\varepsilon+P}-r\mathcal{C}_{r-1}^{(0)} \right.\nonumber\\
     &\left.+\frac{1}{m^2}(r+3)\mathcal{C}^{(0)}_{r+1}
    -\frac{\partial\mathcal{C}^{(0)}_{r-1}}{\partial\ln\beta}+\frac{1}{m^2}\frac{\partial\mathcal{C}^{(0)}_{r+1}}{\partial\ln\beta}\right]\;,\\
    % \tau_{n\Pi}=&\frac{1}{\varepsilon+P}\left[\frac{\beta}{\varepsilon+P}\sum_{r=0,\neq 1}^{N_1}\tau_{0r}^{(1)}J_{r+2,1}-\sum_{r=0,\neq 1}^{N_1}\tau_{0r}^{(1)}r\mathcal{C}_{r-1}^{(0)}\right.\nonumber\\
    % &+\frac{1}{m^2}\sum_{r=0,\neq 1}^{N_1}\tau_{0r}^{(1)}(r+3)\mathcal{C}^{(0)}_{r+1}
    % -\sum_{r=0,\neq 1}^{N_1}\tau_{0r}^{(1)}\frac{\partial\mathcal{C}^{(0)}_{r-1}}{\partial\ln\beta}\nonumber\\
    % &\left.+\frac{1}{m^2}\sum_{r=0,\neq 1}^{N_1}\tau_{0r}^{(1)}\frac{\partial\mathcal{C}^{(0)}_{r+1}}{\partial\ln\beta}\right],\\
    %
     \tau_{n\pi}=&\sum_{r=0,\neq 1}^{N_1}\frac{\tau_{0r}^{(1)}}{\varepsilon+P}\left(\frac{\beta J_{r+2,1}}{\varepsilon+P}-r\mathcal{C}^{(2)}_{r-1}
     -\frac{\partial\mathcal{C}_{r-1}^{(2)}}{\partial\ln\beta}\right)\;,\\
    % \tau_{n\pi}=&\frac{1}{\varepsilon+P}\left[\frac{\beta}{\varepsilon+P}\sum_{r=0,\neq 1}^{N_1}\tau_{0r}^{(1)}J_{r+2,1}-\sum_{r=0,\neq 1}^{N_1}\tau_{0r}^{(1)}r\mathcal{C}^{(2)}_{r-1}\right.\nonumber\\
    % &\left.-\sum_{r=0,\neq 1}^{N_1}\tau_{0r}^{(1)}\frac{\partial\mathcal{C}_{r-1}^{(2)}}{\partial\ln\beta}\right],\\
    %
    \lambda_{nn}=& \sum_{r=0,\neq 1}^{N_1} \frac{\tau_{0r}^{(1)}}{5} \left[(2r+3)\mathcal{C}^{(1)}_r - 2 m^2 (r-1)\mathcal{C}^{(1)}_{r-2}\right]\;,\\
    \lambda_{n\Pi}=&\sum_{r=0,\neq 1}^{N_1}\tau_{0r}^{(1)}\left[\left(\frac{\partial\mathcal{C}^{(0)}_{r-1}}{\partial\alpha}+\frac{1}{h}\frac{\partial\mathcal{C}^{(0)}_{r-1}}{\partial\beta}\right)\right.\\
    &\left.-\frac{1}{m^2}\left(\frac{\partial\mathcal{C}^{(0)}_{r+1}}{\partial\alpha}+\frac{1}{h}\frac{\partial\mathcal{C}^{(0)}_{r+1}}{\partial\beta}\right)\right]\;,\\
    \lambda_{n\pi}=&\sum_{r=0,\neq 1}^{N_1}\tau_{0r}^{(1)}\left(\frac{\partial\mathcal{C}^{(2)}_{r-1}}{\partial\alpha}+\frac{1}{h}\frac{\partial\mathcal{C}^{(2)}_{r-1}}{\partial\beta}\right)\;.
\end{align}

Finally, for the shear-stress tensor we have:
\begin{align}
    \delta_{\pi\pi}=&\sum_{r=0}^{N_2}\tau^{(2)}_{0r}\left[\frac{r+4}{3}\mathcal{C}^{(2)}_r + \mathcal{H}\frac{\partial\mathcal{C}^{(2)}_r}{\partial\alpha}+\bar{\mathcal{H}}\frac{\partial\mathcal{C}^{(2)}_r}{\partial\beta} \right.\nonumber\\
    &\left.-\frac{m^2}{3}(r-1)\mathcal{C}^{(2)}_{r-2}\right]\;,\\
    \tau_{\pi\pi}=&\frac{2}{7}\sum_{r=0}^{N_2} \tau^{(2)}_{0r} \left[(2r+5)\mathcal{C}^{(2)}_{r}-2m^2(r-1)\mathcal{C}^{(2)}_{r-2}\right]\;,\\
    \lambda_{\pi\Pi}=&-\frac{2}{5m^2}\sum_{r=0}^{N_2}\tau^{(2)}_{0r}\left[(r+4)\mathcal{C}^{(0)}_{r+2}-m^2(2r+3)\mathcal{C}^{(0)}_r\right.\nonumber\\
    &\left.+m^4(r-1)\mathcal{C}^{(0)}_{r-2}\right]\;,\\
    \tau_{\pi n}=&\frac{2}{5(\varepsilon+P)}\sum_{r=0}^{N_2}\tau^{(2)}_{0r}\left[(r+5)\mathcal{C}^{(1)}_{r+1}-m^2r\mathcal{C}^{(1)}_{r-1}\right.\nonumber\\
    &\left.+\frac{\partial\mathcal{C}^{(1)}_{r+1}}{\partial\ln\beta}-m^2\frac{\partial\mathcal{C}^{(1)}_{r-1}}{\partial\ln\beta}\right]\;,\\
    \ell_{\pi n} =& \frac{2}{5}\sum_{r=0}^{N_2}\tau^{(2)}_{0r}\left(\mathcal{C}^{(1)}_{r+1}-m^2\mathcal{C}^{(1)}_{r-1}\right)\;,\\
    \lambda_{\pi n}=&\frac{2}{5}\sum_{r=0}^{N_2}\tau^{(2)}_{0r}\left[\left(\frac{\partial\mathcal{C}^{(1)}_{r+1}}{\partial\alpha}+\frac{1}{h}\frac{\partial\mathcal{C}^{(1)}_{r+1}}{\partial\beta}\right)\right.\nonumber\\
    &\left.-m^2\left(\frac{\partial\mathcal{C}^{(1)}_{r-1}}{\partial\alpha}+\frac{1}{h}\frac{\partial\mathcal{C}^{(1)}_{r-1}}{\partial\beta}\right)\right]\;.
    \end{align}

\bibliographystyle{unsrtnat}
\bibliography{bib_IReD}
\end{document}